\documentclass[twocolumn,showpacs,amsmath,prl]{revtex4-1}
\usepackage{graphicx}
\usepackage{subfigure}
\usepackage{bm}
\usepackage{lineno}

\newcommand{\al}{\alpha}
\newcommand{\be}{\beta}

\newcommand{\br}{{\bf r}}

\newcommand{\bk}{{\bf k}}

\newcommand{\bn}{{\bf n}}

\newcommand{\bp}{{\bf p}}

\newcommand{\bq}{{\bf q}}

\newcommand{\bK}{{\bf K}}

\newcommand{\si}{{\sigma}}

\newcommand{\eps}{\epsilon}

\DeclareMathAlphabet{\mathpzc}{OT1}{pzc}{m}{it} 
\begin{document}
\title{Superconductivity in a two-dimensional repulsive Rashba gas at low electron density}
\author{Luyang Wang}
\affiliation{Institute for Advanced Study, Tsinghua University, Beijing, 100084, China}

\date{\today}
\begin{abstract}
We study the superconducting instability and the resulting superconducting states in a two-dimensional repulsive Fermi gas with Rashba spin-orbit coupling at low electron density (namely the Fermi energy $E_F$ is lower than the energy $E_R$ of the Dirac point induced by  Rashba coupling). We find that superconductivity is enhanced as the dimensionless Fermi energy $\epsilon_F$ ($\epsilon_F\equiv E_F/E_R$) decreases, due to the following two reasons. First, the density of states at $\eps_F$ increases as $1/\sqrt{\eps_F}$. Second, the particle-hole bubble becomes more anisotropic, resulting in an increasing effective attraction. The superconducting state is always in the total angular momentum $j_z=+2$ (or $j_z=-2$) channel with Chern number $C=4$ (or $C=-4$), breaking time reversal symmetry spontaneously. Although a putative Leggett mode is expected due to the two-gap nature of the superconductivity, we find that it is always damped. More importantly, once a sufficiently large Zeeman coupling is applied to the superconducting state, the Chern number can be tuned to be $\pm1$ and Majorana zero modes exist in the vortex cores.
\end{abstract}
\maketitle

Despite an effect originating from relativity, spin-orbit coupling (SOC) has found its way into nonrelativistic physics. In condensed matter physics, novel systems with SOC playing a significant role are found recently, such as topological insulators\cite{HasanKane,QiZhang}, two-dimensional (2D) Rashba gases at interfaces of oxides\cite{Ohmoto2004,Triscone2011}, Weyl semimetals\cite{Wan2011} and SOC-induced Mott insulators\cite{BJKim2008} and other states in 5$d$ series\cite{Witczak-Krempa2014}; while in ultracold quantum gases, although atoms are neutral, synthetic SOC can be generated by atom-light interaction (see Ref.\cite{Zhai2015,DalibardRMP} for review). Turning to superconductivity, non-centrosymmetric superconductors, where SOC mixes spin singlet and triplet pairings, have been extensively studied\cite{NCSC,Smidman2016}; and in 2D, superconductivity related to SOC was observed at oxide interfaces\cite{Caviglia2010, Triscone2011}.

\begin{figure}[tb]
\subfigure[]{\includegraphics[width=0.28\textwidth]{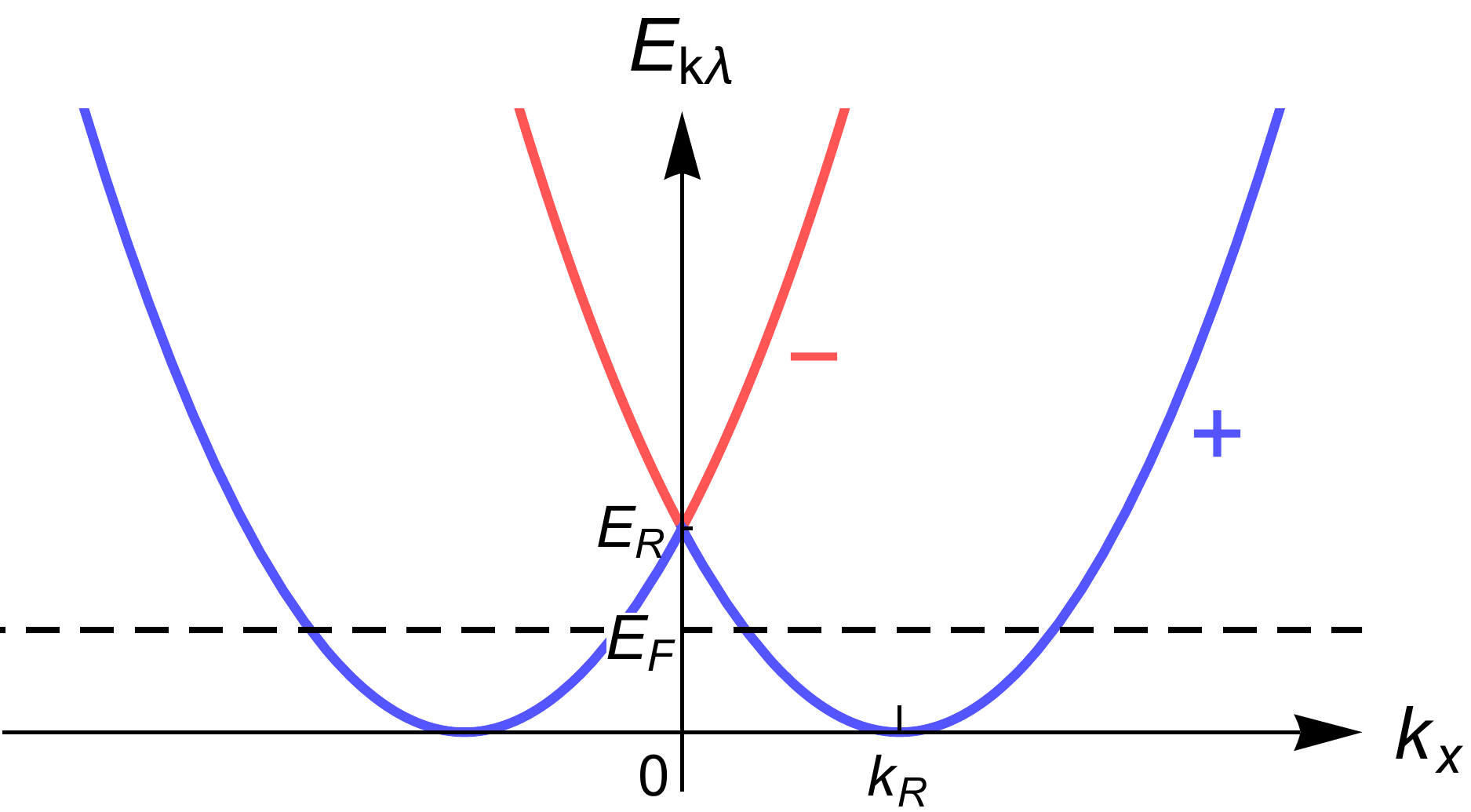}\label{Ek}}
\subfigure[]{\includegraphics[width=0.16\textwidth]{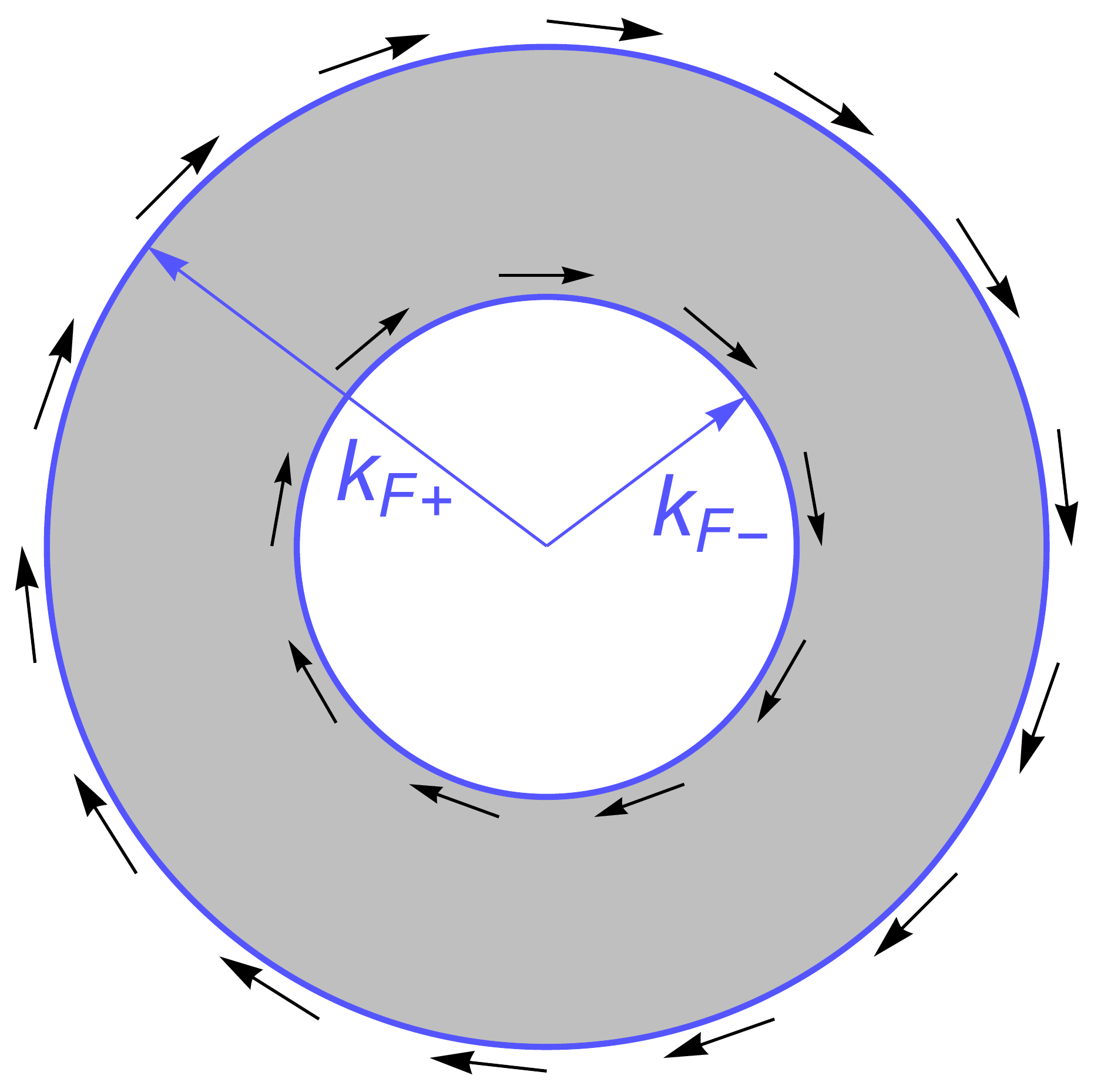}\label{FS}}
\caption{(a) The dispersion relation (fixing $k_y=0$), with the helicity labeled. At low density, the Fermi level indicated by the dashed line is below the Dirac point at energy $E_R$. (b) The two Fermi surfaces, the annulus between which is filled by electrons with helicity +1. The black arrows represent the direction of the spins around the Fermi surfaces.}
\end{figure}

Here, we study a 2D repulsive gas with Rashba SOC at low density. The single-particle Hamiltonian is
\begin{eqnarray}
H=\frac{k^2}{2m}+\al_R(\vec{\sigma}\times\bk)\cdot\hat{\bn},
\end{eqnarray}
where $m$ is the effective mass, $\al_R$ characterizes the strength of Rashba SOC, $\vec{\sigma}$'s components are Pauli matrices, and $\hat{\bn}$ is the direction normal to the 2D system.
By a unitary transformation to helicity basis, one finds the dispersion
\begin{eqnarray}
E_{\bk\lambda}=\frac{(k-\lambda k_R)^2}{2m},
\end{eqnarray}
where $\lambda=\pm1$ is the helicity and $k_R=m\al_R$ is the Rashba momentum. (We have shifted the energy by $k_R^2/(2m)$, which will be compensated by the shift of the Fermi energy.) The spin degeneracy is lifted, resulting in two bands touching at a Dirac point. In this system, the competition between the three energy scales - the Fermi energy $E_F$, Coulomb repulsion and the ``Rashba energy" $E_R=k_R^2/(2m)$ - determines the system's phases. We define the dimensionless Fermi energy by $\eps_F=E_F/E_R$. By ``low density", we refer to the regime $0<\eps_F<1$, as shown in Fig.\ref{Ek}. While in the low-density limit, the Fermi energy is much smaller than Coulomb interaction and SOC energy, and Wigner crystalline phases are found\cite{Berg2012}, we focus on the weak-coupling limit $uN_{tot}\ll1$ (although $N_{tot}\sim1/\sqrt{\eps_F}$ as discussed below), where $u$ is the short-range repulsion and $N_{tot}$ is the total density of states (DOS) at Fermi energy. We investigate the instability of the system to superconducting states. In earlier papers\cite{VafekWang,WangVafek}, the ``high density" case with $\eps_F>1$ was studied. It was found that as $\eps_F$ decreases from a large value to 1, in general, superconductivity is enhanced, and the total angular momentum channel $j_z$ in which Cooper pairs condense decreases from a large value as an arithmetic sequence with a step of 2. Finally, at $\eps_F\gtrsim1$, $j_z$ becomes 2. The superconductivity predominantly resides on the outer Fermi surface, and interband coupling induces a small gap on the inner Fermi surface. The superconducting state breaks time reversal symmetry (TRS), and both Fermi surfaces are fully gapped.

Now, at low density, as shown in Fig.\ref{FS}, there is an electron-like and a hole-like Fermi surface, between which the annulus is filled by electrons. The Fermi wave vectors (rescaled by $k_R$) are $k_{F\mu}=1+\mu\sqrt{\eps_F}$, where $\mu=1$ or $-1$, corresponding to the outer and inner Fermi surface, respectively. Note that the two Fermi surfaces have the same helicity $\lambda=1$. The DOS at the Fermi surface is $N_\mu=\nu_0(\frac{1}{\sqrt{\eps_F}}+\mu)$, where $\nu_0=m/(2\pi)$. It is likely that superconductivity is further enhanced, since the total DOS increases with decreasing $\eps_F$ as $N_{tot}\sim1/\sqrt{\eps_F}$. In addition, we wonder what a topological change of the Fermi surfaces may bring about to the dielectric functions, which can give rise to the attraction in nonzero $j_z$ channels. The enhancement of conventional superconductivity in attractive Rashba gases at low density has been studied in Ref.\cite{Cappelluti2007}, while here we address the problem with repulsion, which leads to unconventional superconductivity.

{\it Renormalization group approach. }We study the system with onsite repulsive interactions, the interacting Hamiltonian of which is
\begin{eqnarray}
H_{int}=\frac{u}{2}\sum_{\bk_1...\bk_4}\sum_{\si\si'}\delta_{\bk_1+\bk_2,\bk_3+\bk_4}c^\dagger_{\bk_1\si}c^\dagger_{\bk_2\si'} c_{\bk_3\si'}c_{\bk_4\si}
\end{eqnarray}
where $u$ is positive. Before proceeding, we need to clarify the parameter regime we are studying - the weak-coupling limit. The potential energy per particle is
\begin{eqnarray}
\overline{E}_{pot}=\frac{1}{2n\Omega}\int d\br d\br'u\delta(\br-\br')n(\br)n(\br')=\frac{1}{2}un
\end{eqnarray}
where $n(\br)\equiv n$ is the uniform density of the gas. The average kinetic energy per particle is
\begin{eqnarray}
\overline{E}_{kin}=\frac{1}{n}\int_0^{E_F}EN(E)dE=\frac{1}{3}E_F
\end{eqnarray}
where the Fermi energy can be expressed as $E_F=(n^2\pi^2)/(4m^2E_R)$. In the weak-coupling limit, $\overline{E}_{pot}\ll\overline{E}_{kin}$, implying $un\ll E_F$, or equivalently $uN_{tot}\ll 1$. As $E_F$ approaches 0, $N_{tot}$ diverges, but we still keep $uN_{tot}\ll 1$.

The interacting Hamiltonian in the helicity basis reads\cite{VafekWang,WangVafek}
\begin{eqnarray}
H_{int}=\frac{u}{16}\sum_{\bk_1...\bk_4}\sum_{\mu\nu\lambda\rho}\delta_{\bk_1+\bk_2,\bk_3+\bk_4} (\mu e^{-i\theta_{\bk_1}}-\nu e^{-i\theta_{\bk_2}})\nonumber\\
\times(\rho e^{i\theta_{\bk_4}}-\lambda e^{i\theta_{\bk_3}}) a^\dagger_{\bk_1\mu}a^\dagger_{\bk_2\nu}a_{\bk_3\lambda}a_{\bk_4\rho}
\end{eqnarray}
where $a_{\bk\mu}$ is the annihilation operator in the helicity basis and $\theta_{\bk}$ is the angle of $\bk$. Following the weak-coupling renormalization group (RG) approach developed in \cite{VafekWang,WangVafek} and integrating out high energy modes from the bandwidth $A$ to a low-energy cutoff $\Omega$, we derive the effective action for the low energy modes
\begin{eqnarray}
S'_{int}&=&\int_0^\be d\tau\sum_{\bk\bk'\mu\lambda}e^{i\phi}\sum_{j_z}e^{ij_z\phi}\nonumber\\
&\times&V_{\mu\lambda}^{r(j_z)}a^*_{\bk\mu}(\tau)a^*_{-\bk\mu}(\tau)a_{-\bk'\lambda(\tau)}a_{\bk'\lambda}(\tau),\label{Eq:Sint}
\end{eqnarray}
where $\be=1/(k_BT)$ and $a$'s and $a^*$'s are Grassmann numbers. We have focused on the Cooper channel, the couplings of which are the only (marginally) relevant ones\cite{Shankar}, and decomposed the couplings into angular momentum channels, where $\phi=\theta_{\bk'}-\theta_{\bk}$.

\begin{figure}[tb]
  \subfigure[]{\includegraphics[width=0.12\textwidth]{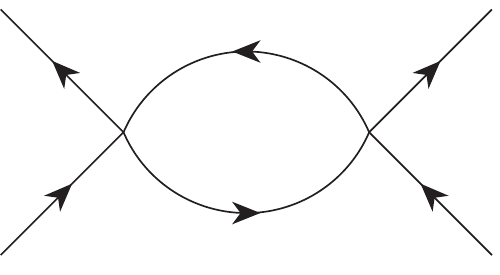}\label{Fig:ph}}\ \ \ \ \ \ \ \ \ \
  \subfigure[]{\includegraphics[width=0.12\textwidth]{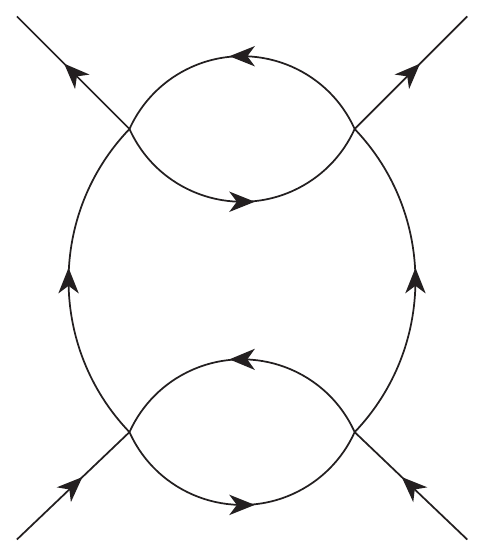}\label{Fig:4th}}
  \caption{(a) The particle-hole bubble. (b) The 4th order correction with logarithmic divergence and finite value in high angular momentum channels.}\label{Fig:FeynmanDiagrams}
\end{figure}

\begin{figure}[tb]
  \begin{tabular}{cc}
  \includegraphics[width=0.23\textwidth]{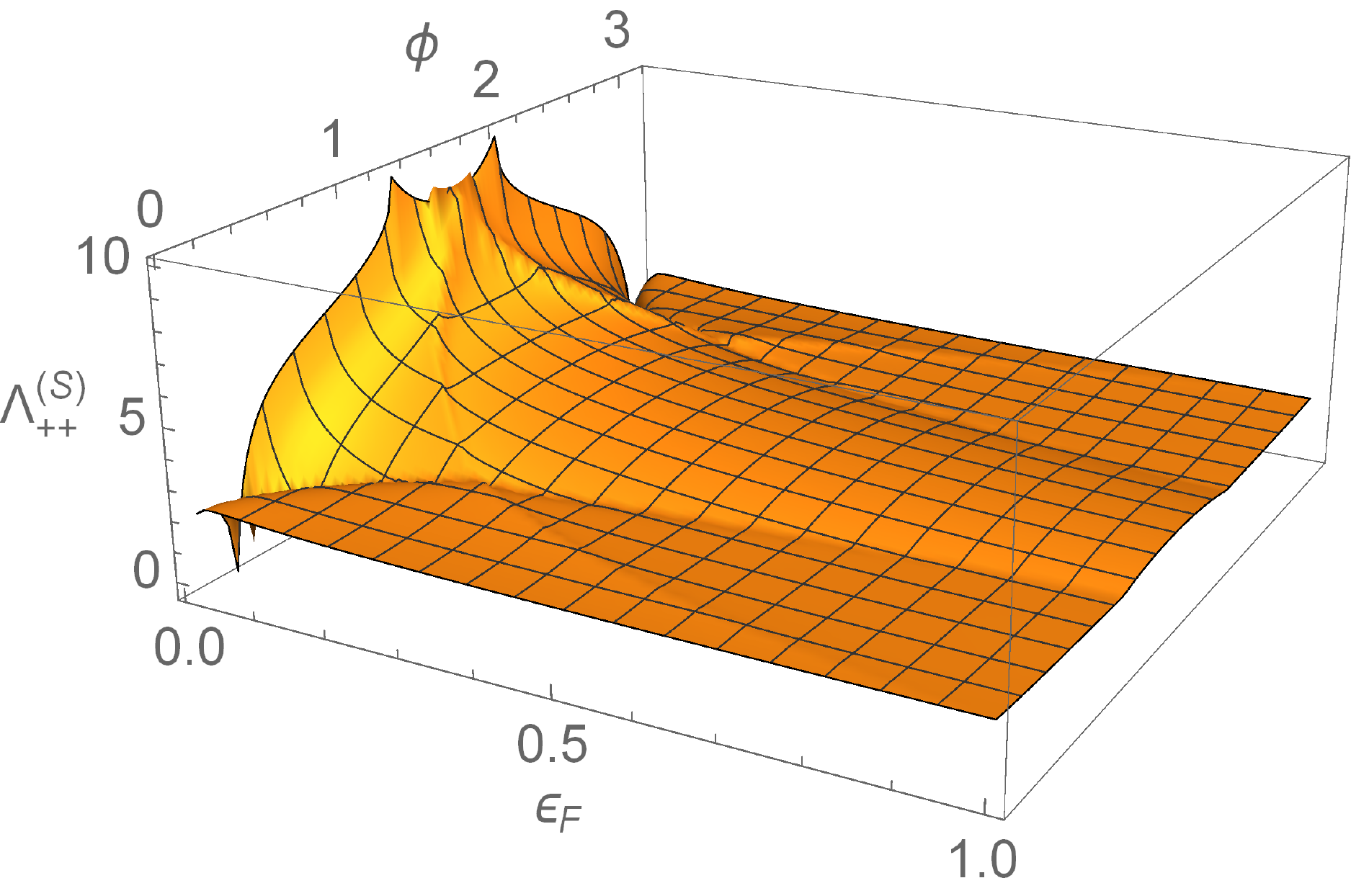}&\includegraphics[width=0.23\textwidth]{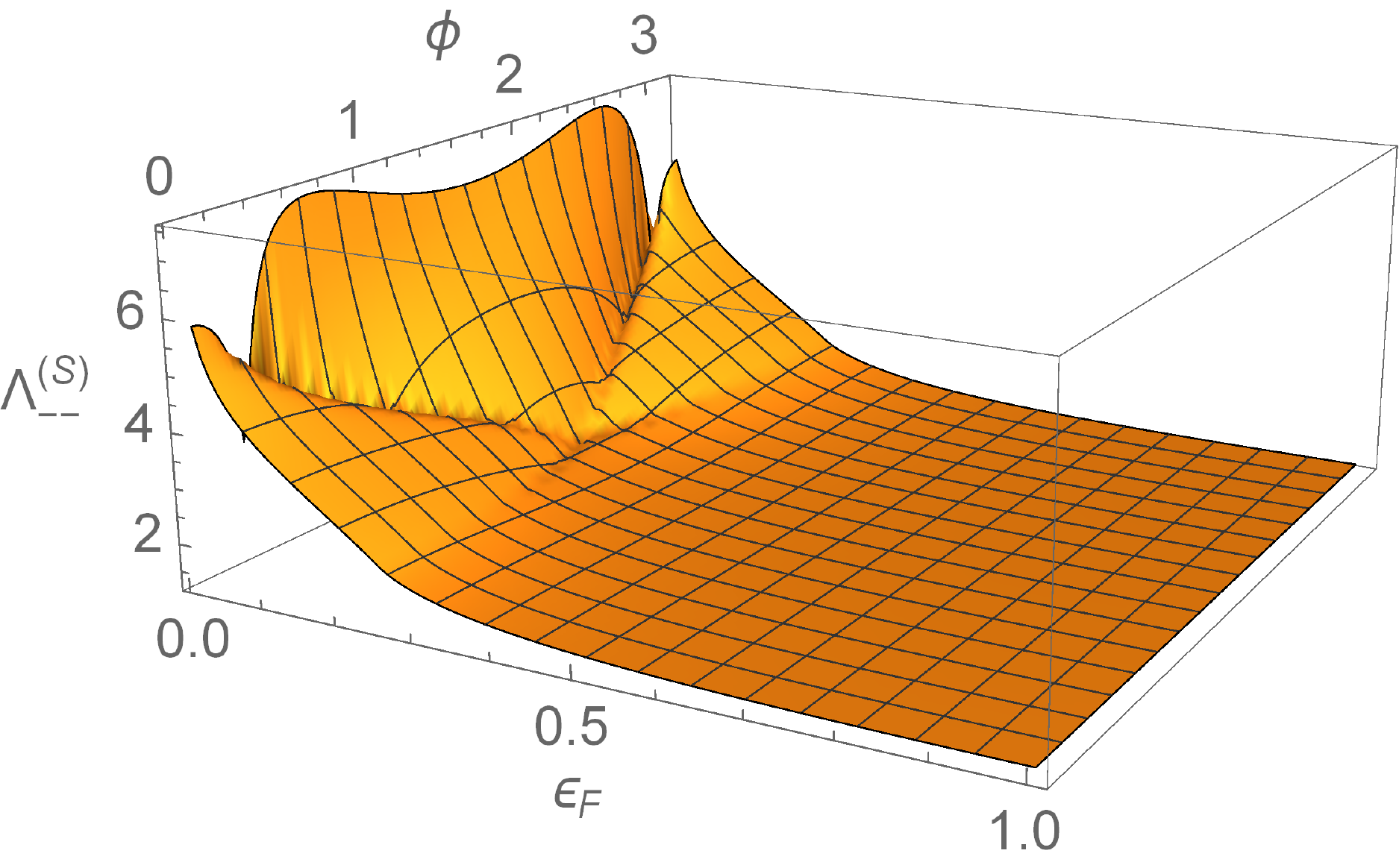}\\
  \includegraphics[width=0.25\textwidth]{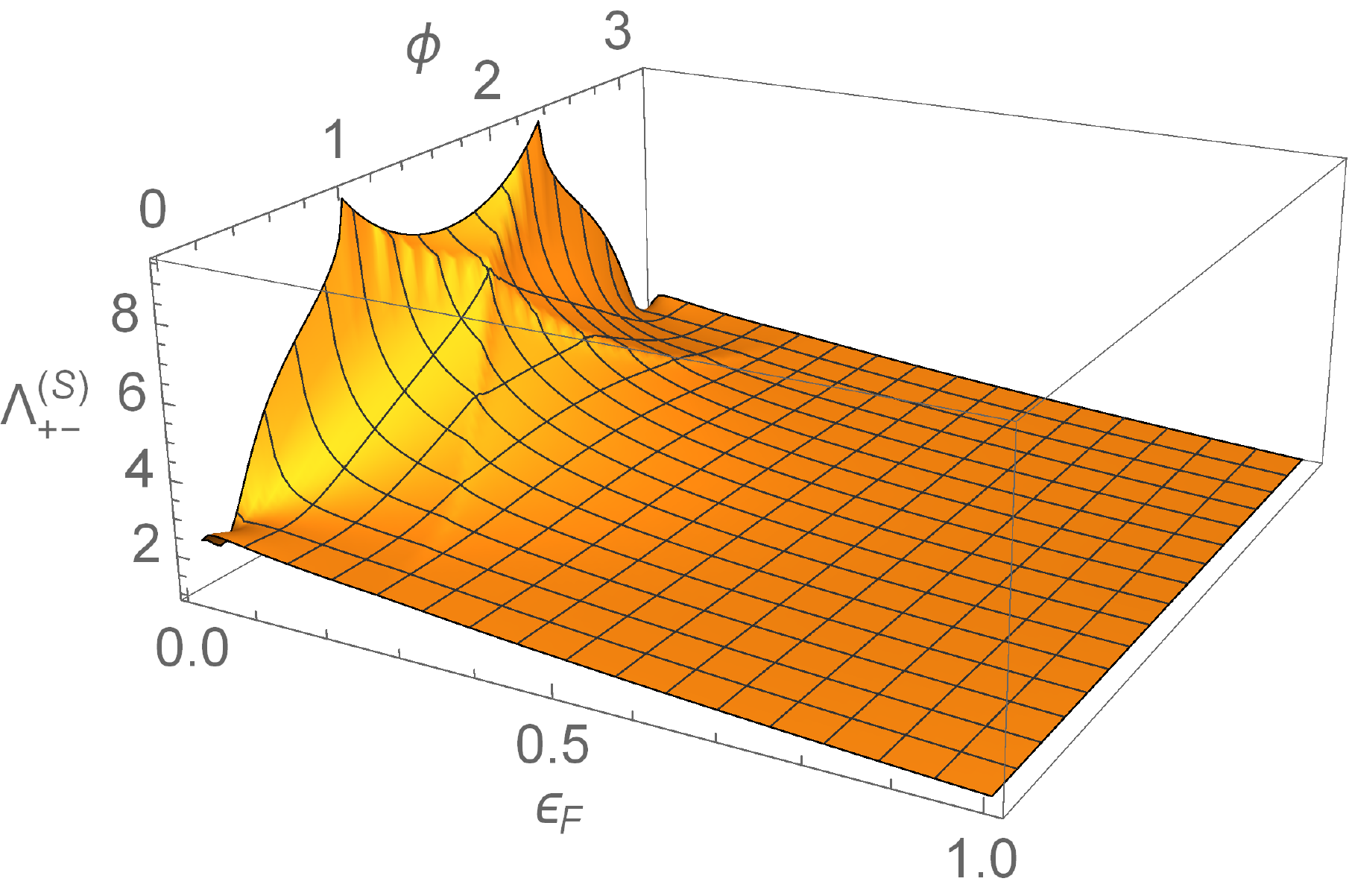}&
  \end{tabular}
  \caption{$\Lambda_{\mu\lambda}^{(S)}$ as a function of $\eps_F$ and $\phi$.}\label{Fig:Pi}
\end{figure}

Since in $j_z=0$ channel, $u$ dominates, one cannot get attractive interactions. Therefore, we go to higher orders and look for nonvanishing terms in higher angular momentum channels. At second order, we have particle-hole bubble and particle-particle bubble, the latter of which only has $j_z=0$ component. The correction from the particle-hole bubble (shown in Fig.\ref{Fig:ph}) is $\frac{u^2}{2^6}(\Pi(\bk,\bk')-\Pi(-\bk,\bk'))$, where the dielectric function is
\begin{eqnarray}
\Pi(\bk,\bk')&=&\sum_{\al\be}\int\frac{d^2\bp}{(2\pi)^2}\frac{n_F(E_{\bp\al})-n_F(E_{\bp+\bk-\bk'\be})} {E_{\bp\al}-E_{\bp+\bk-\bk'\be}}\nonumber\\
&\times&F_{\al\be}(\bk,\bk',\bp),
\end{eqnarray}
where
\begin{eqnarray}
F_{\al\be}(\bk,\bk',\bp)&=&(\al e^{-i\theta_\bp}-e^{-i\theta_\bk})(e^{i\theta_{\bk'}}-\be e^{i\theta_{\bp-\bk'+\bk}})\nonumber\\
&\times&(\be e^{-i\theta_{\bp-\bk'+\bk}}-e^{-i\theta_{-\bk}})(e^{i\theta_{-\bk'}}-\al e^{i\theta_\bp}).\nonumber
\end{eqnarray}
Since Cooper pairs are expected to form between electrons near the Fermi surfaces, $\bk$ and $\bk'$ are restricted to be at Fermi surface $\mu$ and $\lambda$, respectively. Straightforward calculations show that $\Pi(\bk,\bk')$ can be written in the form
\begin{eqnarray}
\Pi(\bk,\bk')=e^{i\phi}2m\Lambda_{\mu\lambda}(\eps_F,\cos\phi),
\end{eqnarray}
where $\Lambda_{\mu\lambda}(\eps_F,\cos\phi)$ is a real function that depends on the dimensionless Fermi energy $\eps_F$, but not on $E_F$ and $E_R$ independently. Then the renormalized coupling appearing in Eq.(\ref{Eq:Sint}) reads
\begin{eqnarray}
V_{\mu\lambda}^{r(j_z)}=\frac{u^2m}{2^5}V_{\mu\lambda}^{(j_z)}+...,
\end{eqnarray}
where $V_{\mu\lambda}^{(j_z)}$ is the $j_z$-th Fourier component of $\Lambda_{\mu\lambda}^{(S)}(\eps_F,\cos\phi)\equiv\frac{1}{2}(\Lambda_{\mu\lambda}(\eps_F,\cos\phi)+\Lambda_{\mu\lambda}(\eps_F,-\cos\phi))$. The functions $\Lambda_{\mu\lambda}^{(S)}(\eps_F,\cos\phi)$ are plotted in Fig.\ref{Fig:Pi}. At $\eps_F\rightarrow1^-$, $\Lambda_{++}^{(S)}$ and $\Lambda_{--}^{(S)}$ connect with the same functions at $\eps_F\rightarrow1^+$ calculated in Ref.\cite{VafekWang,WangVafek}, but $\Lambda_{+-}^{(S)}$ changes sign due to the change of the helicity of the inner Fermi surface. Clearly, the functions depend more strongly on $\phi$ at smaller $\eps_F$.

\begin{figure}[tb]
  \includegraphics[width=0.4\textwidth]{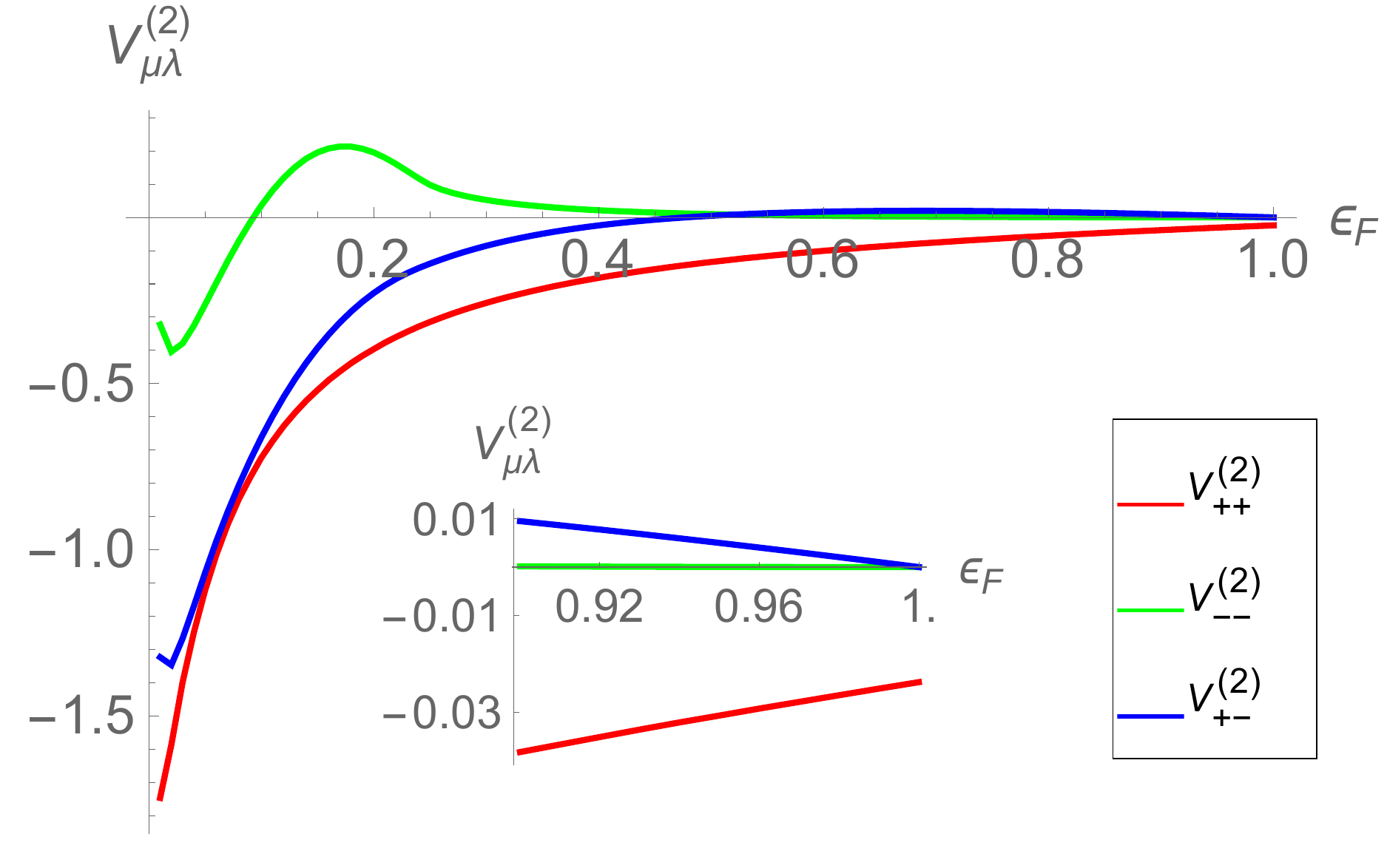}
  \caption{The intraband and interband couplings in $j_{z0}$-channel as a function of $\eps_F$. At $\eps_F\rightarrow1^-$, $V_{--}^{(j_{z0})}$ and $V_{+-}^{(j_{z0})}$ vanish.}\label{Fig:V}
\end{figure}

\begin{figure}[tb]
  \includegraphics[width=0.4\textwidth]{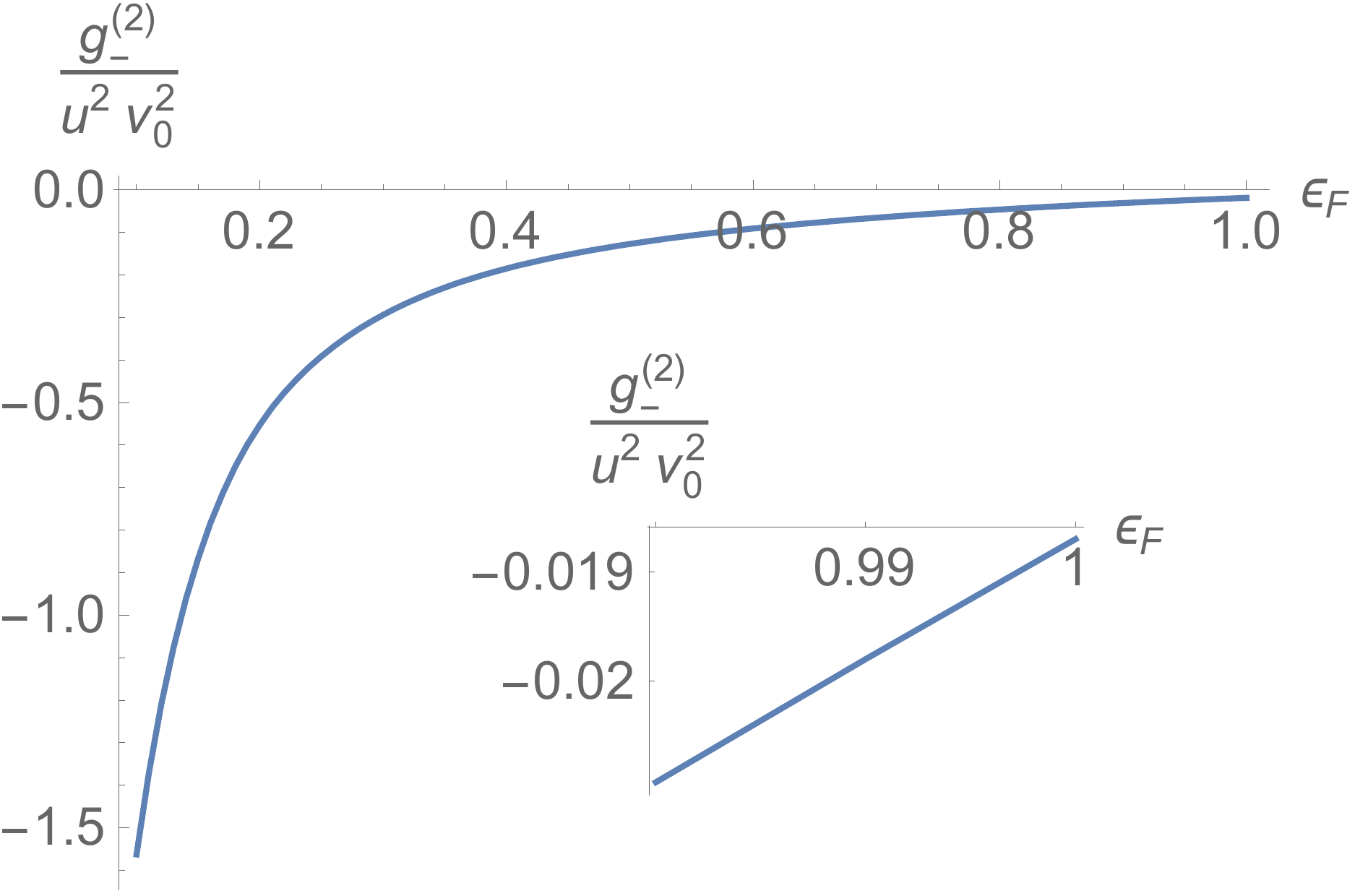}
  \caption{The effective coupling in units of $u^2\nu_0^2$. At $\eps_F\rightarrow1^-$, it goes to $-0.0187$ that agrees with the value at $\eps_F\rightarrow 1^+$ derived in Ref.\cite{VafekWang,WangVafek}.}\label{Fig:g1}
\end{figure}

\begin{figure}[tb]
  \includegraphics[width=0.4\textwidth]{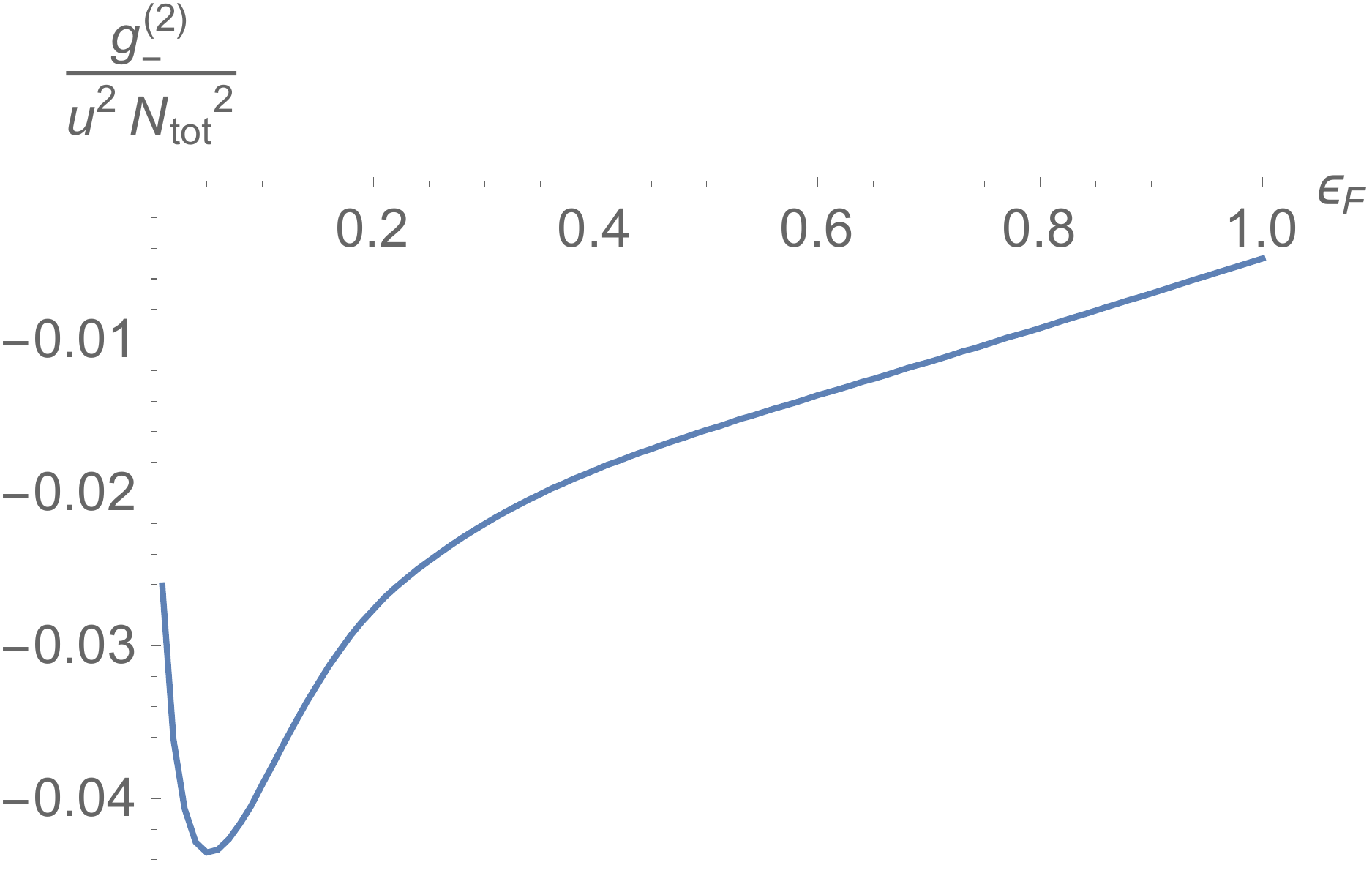}
  \caption{The effective coupling in units of $u^2 N_{tot}^2$.}\label{Fig:g2}
\end{figure}

Up to the fourth order of $u$, there is only one term that satisfies two conditions: (i) being finite in nonzero angular momentum channel; and (ii) having a logarithmic divergence $\ln(A/\Omega)$, which may give rise to an instability. This term is shown in Fig.\ref{Fig:4th}. Including it, the renormalized couplings become
\begin{eqnarray}
V_{\mu\lambda}^{r(j_z)}=\frac{u^2m}{2^5}V_{\mu\lambda}^{(j_z)}-\frac{u^4m^2}{2^9}\sum_{\al}N_\al V_{\mu\al}^{(j_z)}V_{\al\lambda} ^{(j_z)}\ln\frac{A}{\Omega}+...
\end{eqnarray}
Defining the dimensionless bare coupling $g_{\mu\lambda}^{(j_z)}=\frac{u^2m}{16}\sqrt{N_\mu N_\lambda}V_{\mu\lambda}^{(j_z)}$ and the dimensionless renormalized coupling $g^{r(j_z)}_{\mu\lambda}=2\sqrt{N_\mu N_\lambda}V^{r(j_z)}_{\mu\lambda}$, we find the RG flow equation
\begin{eqnarray}
\frac{dg^{r(j_z)}}{d\ln\Omega}=g^{r(j_z)}*g^{r(j_z)}
\end{eqnarray}
where ``$*$" is the matrix multiplication, and the bare couplings have been replaced by the renormalized couplings. The solution is
\begin{eqnarray}
g_{\pm}^{r(j_z)}(\Omega)=\frac{1}{1/g_{\pm}^{(j_z)}+\ln(A/\Omega)},
\end{eqnarray}
with the initial condition
\begin{eqnarray}
g_{\pm}^{(j_z)}&=&\frac{u^2m}{16}\left(\frac{1}{2}(N_+V_{++}^{(j_z)} +N_-V_{--}^{(j_z)})\right.\nonumber\\
&\pm&\left.\sqrt{\frac{1}{4}(N_+V_{++}^{(j_z)} -N_-V_{--}^{(j_z)})^2+N_+N_-{V_{+-}^{(j_z)}}^2}\right).
\end{eqnarray}
The scale of the superconducting transition temperature is given by the largest energy at which the renormalized coupling diverges
\begin{eqnarray}
T_c\sim\Omega^{*(j_{z0})}=Ae^{-\left|\frac{1}{g_-^{(j_{z0})}}\right|},
\end{eqnarray}
where $j_{z0}$ is chosen in such a way that, at a given $\eps_F$, $g_-^{(j_{z0})}$ is the most negative among all the $g_{\pm}^{(j_z)}$'s. We find that as long as $\eps_F<1$, $|j_{z0}|\equiv2$, although it increases in general with $\eps_F$ when $\eps_F>1$\cite{VafekWang,WangVafek}. The interband coupling $V_{+-}^{(j_z)}$ and intraband couplings $V_{++}^{(j_z)}$ and $V_{--}^{(j_z)}$ are shown in Fig.\ref{Fig:V}. As can be seen, $V_{--}$ and $V_{+-}$ are negligible above $\eps_F\sim0.5$, and superconductivity predominantly resides on the outer Fermi surface; and the inner Fermi surface participates only at quite low density. The effective coupling $g_-^{(j_{z0})}$ is plotted in Fig.\ref{Fig:g1} and Fig.\ref{Fig:g2} in units of $u^2\nu_0^2$ and $u^2 N_{tot}^2$, respectively.


Although it seems $|g_-^{(j_{z0})}|$ can acquire any large value from Fig.\ref{Fig:g1}, in our weak-coupling RG approach, the result is not justified for $uN_{tot}\gtrsim1$, and we see from Fig.\ref{Fig:g2} that the effective coupling is still restricted to be small in the justifiable parameter regime. However, $T_c$ can increase by many orders of magnitude as $\eps_F$ decreases. If $u$ is fixed, a superconductor-insulator transition is expected as the density of the gas decreases since Wigner crystal state should exist at strong coupling\cite{Berg2012}, but we are not concerned with this case.

{\it Topological phase diagram.}
A recent theme in condensed matter physics is to search for Majorana fermions in various systems\cite{Alicea}. In addition to intrinsic chiral $p$-wave superconductors\cite{KopninSalomaa,ReadGreen,Ivanov2001,Tewari2007}, spin-orbit coupled systems, such as topological insulators\cite{FuKane2008}, semiconductors with Zeeman splitting in proximity to $s$-wave superconductors\cite{Sau2010PRL} and hole-doped semiconductors\cite{Mao2011} also support Majorana fermion modes in the vortex cores. We show that the intrinsic superconducting state we have found can also host Majorana fermions once a sufficiently large Zeeman field is applied.

\begin{figure}[tb]
  \subfigure[]{\includegraphics[width=0.2\textwidth]{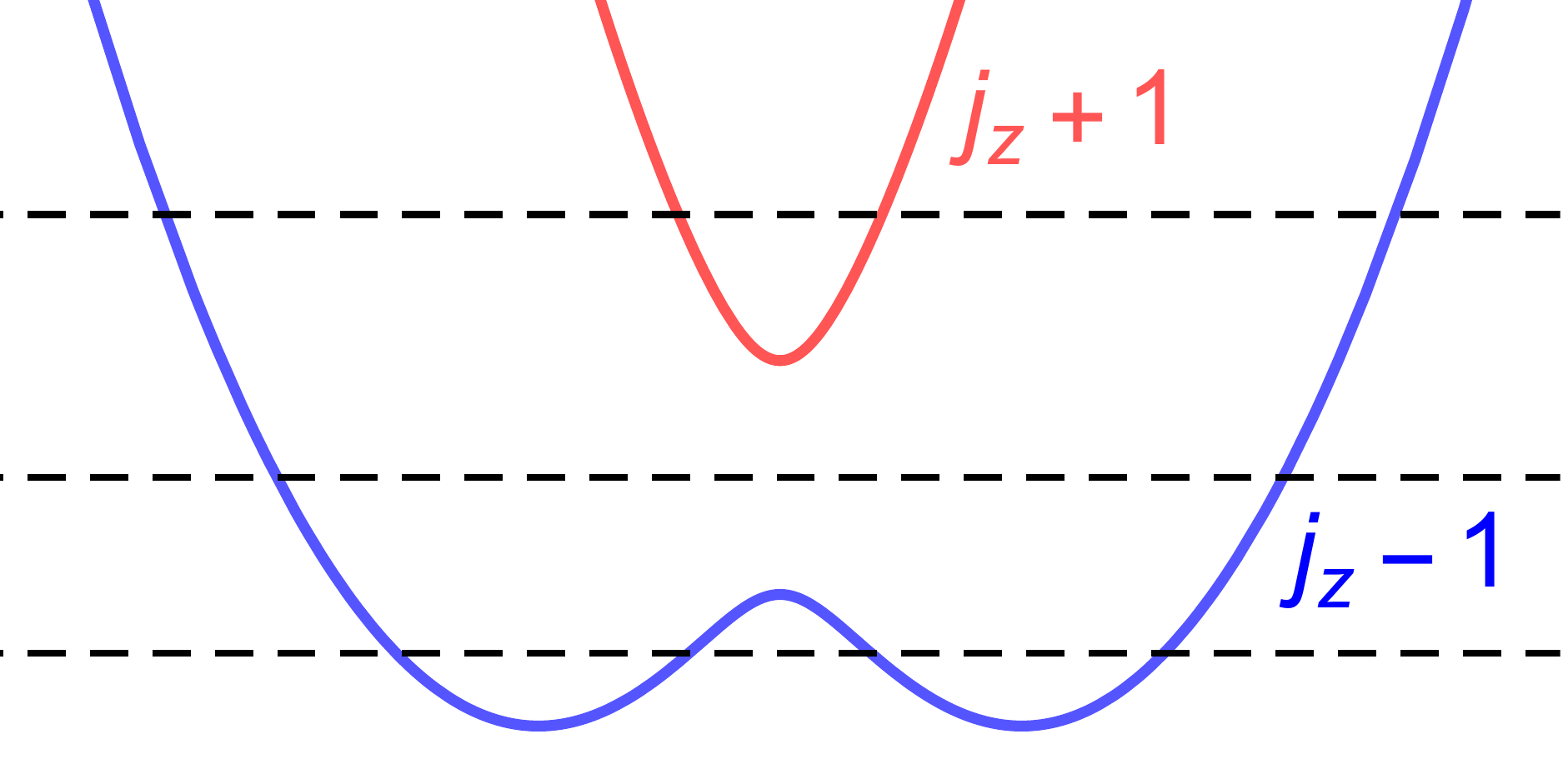}\label{Fig:Zeeman}}\ \
  \subfigure[]{\includegraphics[width=0.25\textwidth]{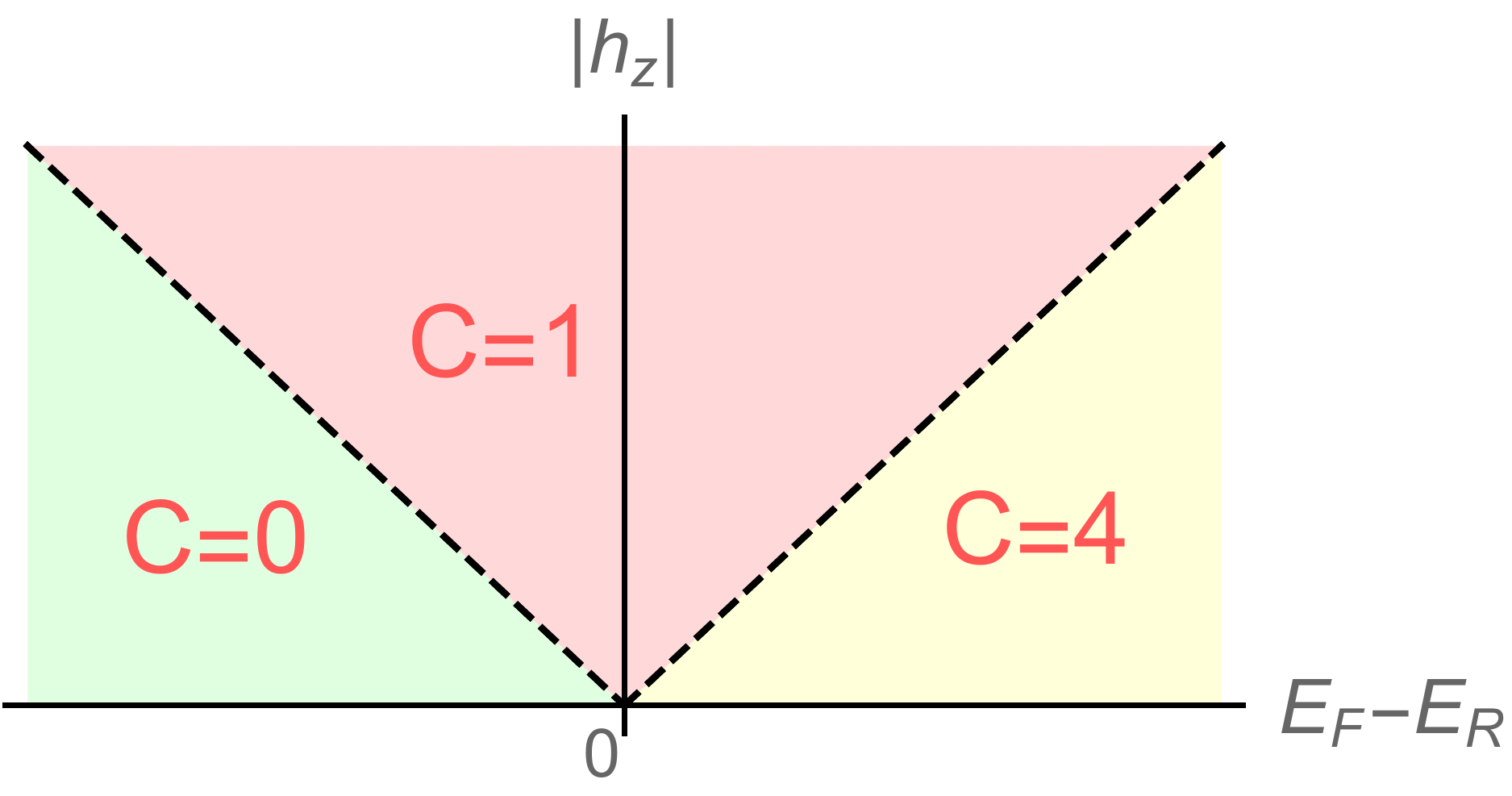}\label{Fig:Chern}}
  \caption{(a) The Dirac point is gapped by the out-of-plane Zeeman field. In the superconducting state, the inner band carries Chern number $j_z+1$, and the outer band carries Chern number $j_z-1$. The three dashed lines indicate the Fermi level in the three corresponding topological phases. (b) Topological phase diagram. Different topological phases are indicated by different colors, with the corresponding Chern number labeled.}\label{Fig:phases}
\end{figure}

As discussed in Ref.\cite{VafekWang,WangVafek}, the ground state breaks TRS spontaneously and the system goes to either $j_z=2$ or $j_z=-2$ superconducting state, both of which have the same energy. Due to SOC, the paring term in the Hamiltonian has both triplet and singlet part, which reads
\begin{eqnarray}
&&\Delta_t e^{i(j_z-1)\theta_\bk}c^\dagger_{\bk\uparrow}c^\dagger_{-\bk\uparrow} +\Delta_te^{i(j_z+1)\theta_\bk}c^\dagger_{\bk\downarrow}c^\dagger_{-\bk\downarrow} \nonumber\\ &+&\Delta_s e^{ij_z\theta_\bk}(c^\dagger_{\bk\uparrow}c^\dagger_{-\bk\downarrow} -c^\dagger_{\bk\downarrow}c^\dagger_{-\bk\uparrow})+h.c.,
\end{eqnarray}
where $\Delta_t$ and $\Delta_s$ are the triplet and singlet pairing strength, respectively, and $\theta_\bk$ is the angle between $\bk$ and $k_x$-axis. To search for Majorana fermions, instead of solving the Bogoliubov-de Gennes equations in the presence of vortices, we apply an index theorem proved in \cite{Roy2010PRL} that superconductors with an odd Chern number can support Majorana zero modes. In Ref.\cite{VafekWang,WangVafek}, the Chern number $C$ of this state has been shown to be $2j_z$ if $E_F>E_R$, and 0 if $E_F<E_R$. To have an odd $C$, we apply a Zeeman field $h_z$\cite{Sau2010PRL} which couples to the system as
\begin{eqnarray}
h_z(c^\dagger_{\bk\uparrow}c_{\bk\uparrow}-c^\dagger_{\bk\downarrow}c_{\bk\downarrow}).
\end{eqnarray}
Since now the Hamiltonian breaks time reversal symmetry, the two states with $j_z=\pm2$ have different energies. As pointed in \cite{Farrell}, if $h_z<0$, the system favors $j_z=2$ state; and if $h_z>0$, it favors $j_z=-2$ state. These two cases are related by time reversal operation, so we just consider the former case. We calculate the Chern number and find three topological phases, depending on the position of Fermi level and $h_z$, as shown in Fig.\ref{Fig:phases}. The topological phase diagram can be explained as follows. When $h_z<0$, the Dirac point is gapped, and Chern number is well defined for each band in the superconducting state. Due to the winding of spin around $\bk=0$, the band with helicity $-1$ ($+1$) carries Chern number 1 ($-1$). In addition, the phase winding of the order parameter superimposes $j_z$ to the Chern number of each band. Therefore, the inner and outer band carry Chern number $j_z+1$ and $j_z-1$, respectively. When $E_F-E_R>|h_z|$, the Fermi level crosses both bands, and the Chern number is the sum of the two, $C=2j_z$. When $E_F-E_R<-|h_z|$, the Fermi level either crosses the outer band twice, in which case the electron pocket and hole pocket contribute opposite Chern number, thus $C=0$, or does not cross any band and hence $C=0$. Between these two parameter regimes, i.e. when $|h_z|>|E_F-E_R|$, the Fermi level only crosses the outer band, and then $C=j_z-1$. In this case, according to the index theorem mentioned above, Majorana zero modes exist at the edge and in the vortex cores. At low density, only the left part of the phase diagram in Fig.\ref{Fig:Chern} is available.

{\it Collective modes.}
Collective modes in superconductors were first studied by Bogoliubov\cite{Bogoliubov} and Anderson\cite{Anderson1958a,Anderson1958b}. They found a Goldstone mode accompanying the spontaneous breaking of $U(1)$ gauge symmetry in a neutral system, which corresponds to the phase oscillations of the superconducting order parameter. In a charged system, this so called Bogoliubov-Anderson-Goldstone (BAG) mode is pushed up to the plasma frequency. In two-band superconductors, Leggett predicted another collective mode corresponding to the oscillations of the relative phase of the two superconducting condensates\cite{Leggett1966}. In the superconducting state derived above, due to the two-gap nature, in addition to BAG mode, Leggett mode may also exist. The detailed calculations are carried out in the supplemental material. Effectively, in $j_z=2$ channel, we have a two-band $p+ip$ superconductor. Actually, if the bare interaction were attractive, superconductivity would occur in $j_z=0$ channel, and the superconducting state would also have a two-band $p+ip$ nature. Thus our approach also applies to that case. We find both BAG and Leggett modes, while the former is pushed to plasma energy, the latter has a dispersion
\begin{eqnarray}
\omega^2=\omega_0^2+v^2K^2,
\end{eqnarray}
where
\begin{eqnarray}
\omega_0^2&=&\frac{N_1+N_2}{2N_1N_2}\frac{8|g_{12}|\Delta_1\Delta_2}{g_{11}g_{22}-g_{12}^2},\\
v^2&=&\frac{(N_1+N_2)c_1^2c_2^2}{N_1c_1^2+N_2c_2^2}.
\end{eqnarray}
In the above, $N_\lambda$'s are the density of states, $g_{\mu\lambda}\sim V_{\mu\lambda}$, and
\begin{eqnarray}\label{Eq:c}
c_\lambda^2=\frac{n_\lambda}{mN_\lambda} +2\frac{\Delta_\lambda^2}{k_{F\lambda}^2}(1-2\ln\frac{A}{\Delta_\lambda}),
\end{eqnarray}
where $n_\lambda$ is the total particle number in band $\lambda$, and $\lambda=1,2$ labels the outer and inner band. Note that the second term in Eq.(\ref{Eq:c}) does not appear in a two-band $s$-wave superconductor. In order for Leggett mode to be undamped, it is necessary for $\omega_0^2$ to be positive, hence $g_{11}g_{22}-g_{12}^2>0$. However, the whole parameter regime shown in Fig.\ref{Fig:V} does not satisfy this condition, thus this mode must be damped in our model. But undamped Leggett mode could exist in other SOC superconductors.\cite{Bittner2015}.

{\it Discussion and conclusion. }The superconductivity in repulsive systems was first studied by Kohn and Luttinger\cite{Kohn1965}. Due to the extremely low transition temperature, to observe such superconductivity is a formidable task. Now we estimate $T_c$ in our setting. Assuming the bandwidth $A$ is of order $\sim eV$, and $uN_{tot}\lesssim 1$, then $T_c$ can achieve $\sim10^{-7}$K. While this is still a too low temperature for experimental observation in condensed matter, it is achievable in ultracold atoms.

In conclusion, we have investigated the superconducting instability of a 2D Rashba gas with repulsive interaction at low density, in the weak-coupling limit. As the density decreases, the superconducting transition temperature increases significantly. The superconducting state is always in $j_z=2$ channel. When a Zeeman field is applied, the state can have an odd Chern number, and hence Majorana zero modes are supported. Although Leggett mode does not exist due to the specific parameters, we expect it to appear in other spin-orbit coupled superconductors because of the two-gap nature.

{\it Acknowledgements}: We would like to thank Hong Yao for stimulating discussions. This work was supported in part by the NSFC under Grant No. 11474175 at Tsinghua University.

\bibliography{SOCbib}

\begin{thebibliography}{34}%
\makeatletter
\providecommand \@ifxundefined [1]{%
 \@ifx{#1\undefined}
}%
\providecommand \@ifnum [1]{%
 \ifnum #1\expandafter \@firstoftwo
 \else \expandafter \@secondoftwo
 \fi
}%
\providecommand \@ifx [1]{%
 \ifx #1\expandafter \@firstoftwo
 \else \expandafter \@secondoftwo
 \fi
}%
\providecommand \natexlab [1]{#1}%
\providecommand \enquote  [1]{``#1''}%
\providecommand \bibnamefont  [1]{#1}%
\providecommand \bibfnamefont [1]{#1}%
\providecommand \citenamefont [1]{#1}%
\providecommand \href@noop [0]{\@secondoftwo}%
\providecommand \href [0]{\begingroup \@sanitize@url \@href}%
\providecommand \@href[1]{\@@startlink{#1}\@@href}%
\providecommand \@@href[1]{\endgroup#1\@@endlink}%
\providecommand \@sanitize@url [0]{\catcode `\\12\catcode `\$12\catcode
  `\&12\catcode `\#12\catcode `\^12\catcode `\_12\catcode `\%12\relax}%
\providecommand \@@startlink[1]{}%
\providecommand \@@endlink[0]{}%
\providecommand \url  [0]{\begingroup\@sanitize@url \@url }%
\providecommand \@url [1]{\endgroup\@href {#1}{\urlprefix }}%
\providecommand \urlprefix  [0]{URL }%
\providecommand \Eprint [0]{\href }%
\providecommand \doibase [0]{http://dx.doi.org/}%
\providecommand \selectlanguage [0]{\@gobble}%
\providecommand \bibinfo  [0]{\@secondoftwo}%
\providecommand \bibfield  [0]{\@secondoftwo}%
\providecommand \translation [1]{[#1]}%
\providecommand \BibitemOpen [0]{}%
\providecommand \bibitemStop [0]{}%
\providecommand \bibitemNoStop [0]{.\EOS\space}%
\providecommand \EOS [0]{\spacefactor3000\relax}%
\providecommand \BibitemShut  [1]{\csname bibitem#1\endcsname}%
\let\auto@bib@innerbib\@empty
\bibitem [{\citenamefont {Hasan}\ and\ \citenamefont {Kane}(2010)}]{HasanKane}%
  \BibitemOpen
  \bibfield  {author} {\bibinfo {author} {\bibfnamefont {M.~Z.}\ \bibnamefont
  {Hasan}}\ and\ \bibinfo {author} {\bibfnamefont {C.~L.}\ \bibnamefont
  {Kane}},\ }\href {\doibase 10.1103/RevModPhys.82.3045} {\bibfield  {journal}
  {\bibinfo  {journal} {Rev. Mod. Phys.}\ }\textbf {\bibinfo {volume} {82}},\
  \bibinfo {pages} {3045} (\bibinfo {year} {2010})}\BibitemShut {NoStop}%
\bibitem [{\citenamefont {Qi}\ and\ \citenamefont {Zhang}(2011)}]{QiZhang}%
  \BibitemOpen
  \bibfield  {author} {\bibinfo {author} {\bibfnamefont {X.-L.}\ \bibnamefont
  {Qi}}\ and\ \bibinfo {author} {\bibfnamefont {S.-C.}\ \bibnamefont {Zhang}},\
  }\href {\doibase 10.1103/RevModPhys.83.1057} {\bibfield  {journal} {\bibinfo
  {journal} {Rev. Mod. Phys.}\ }\textbf {\bibinfo {volume} {83}},\ \bibinfo
  {pages} {1057} (\bibinfo {year} {2011})}\BibitemShut {NoStop}%
\bibitem [{\citenamefont {Ohmoto}\ and\ \citenamefont
  {Hwang}(2004)}]{Ohmoto2004}%
  \BibitemOpen
  \bibfield  {author} {\bibinfo {author} {\bibfnamefont {A.}~\bibnamefont
  {Ohmoto}}\ and\ \bibinfo {author} {\bibfnamefont {H.~Y.}\ \bibnamefont
  {Hwang}},\ }\href@noop {} {\bibfield  {journal} {\bibinfo  {journal}
  {Nature}\ }\textbf {\bibinfo {volume} {427}},\ \bibinfo {pages} {423}
  (\bibinfo {year} {2004})}\BibitemShut {NoStop}%
\bibitem [{\citenamefont {Zubko}\ \emph {et~al.}(2011)\citenamefont {Zubko},
  \citenamefont {Gariglio}, \citenamefont {Gabay}, \citenamefont {Ghosez},\
  and\ \citenamefont {Triscone}}]{Triscone2011}%
  \BibitemOpen
  \bibfield  {author} {\bibinfo {author} {\bibfnamefont {P.}~\bibnamefont
  {Zubko}}, \bibinfo {author} {\bibfnamefont {S.}~\bibnamefont {Gariglio}},
  \bibinfo {author} {\bibfnamefont {M.}~\bibnamefont {Gabay}}, \bibinfo
  {author} {\bibfnamefont {P.}~\bibnamefont {Ghosez}}, \ and\ \bibinfo {author}
  {\bibfnamefont {J.-M.}\ \bibnamefont {Triscone}},\ }\href {\doibase
  10.1146/annurev-conmatphys-062910-140445} {\bibfield  {journal} {\bibinfo
  {journal} {Annual Review of Condensed Matter Physics}\ }\textbf {\bibinfo
  {volume} {2}},\ \bibinfo {pages} {141} (\bibinfo {year} {2011})}\BibitemShut
  {NoStop}%
\bibitem [{\citenamefont {Wan}\ \emph {et~al.}(2011)\citenamefont {Wan},
  \citenamefont {Turner}, \citenamefont {Vishwanath},\ and\ \citenamefont
  {Savrasov}}]{Wan2011}%
  \BibitemOpen
  \bibfield  {author} {\bibinfo {author} {\bibfnamefont {X.}~\bibnamefont
  {Wan}}, \bibinfo {author} {\bibfnamefont {A.~M.}\ \bibnamefont {Turner}},
  \bibinfo {author} {\bibfnamefont {A.}~\bibnamefont {Vishwanath}}, \ and\
  \bibinfo {author} {\bibfnamefont {S.~Y.}\ \bibnamefont {Savrasov}},\ }\href
  {\doibase 10.1103/PhysRevB.83.205101} {\bibfield  {journal} {\bibinfo
  {journal} {Phys. Rev. B}\ }\textbf {\bibinfo {volume} {83}},\ \bibinfo
  {pages} {205101} (\bibinfo {year} {2011})}\BibitemShut {NoStop}%
\bibitem [{\citenamefont {Kim}\ \emph {et~al.}(2008)\citenamefont {Kim},
  \citenamefont {Jin}, \citenamefont {Moon}, \citenamefont {Kim}, \citenamefont
  {Park}, \citenamefont {Leem}, \citenamefont {Yu}, \citenamefont {Noh},
  \citenamefont {Kim}, \citenamefont {Oh}, \citenamefont {Park}, \citenamefont
  {Durairaj}, \citenamefont {Cao},\ and\ \citenamefont
  {Rotenberg}}]{BJKim2008}%
  \BibitemOpen
  \bibfield  {author} {\bibinfo {author} {\bibfnamefont {B.~J.}\ \bibnamefont
  {Kim}}, \bibinfo {author} {\bibfnamefont {H.}~\bibnamefont {Jin}}, \bibinfo
  {author} {\bibfnamefont {S.~J.}\ \bibnamefont {Moon}}, \bibinfo {author}
  {\bibfnamefont {J.-Y.}\ \bibnamefont {Kim}}, \bibinfo {author} {\bibfnamefont
  {B.-G.}\ \bibnamefont {Park}}, \bibinfo {author} {\bibfnamefont {C.~S.}\
  \bibnamefont {Leem}}, \bibinfo {author} {\bibfnamefont {J.}~\bibnamefont
  {Yu}}, \bibinfo {author} {\bibfnamefont {T.~W.}\ \bibnamefont {Noh}},
  \bibinfo {author} {\bibfnamefont {C.}~\bibnamefont {Kim}}, \bibinfo {author}
  {\bibfnamefont {S.-J.}\ \bibnamefont {Oh}}, \bibinfo {author} {\bibfnamefont
  {J.-H.}\ \bibnamefont {Park}}, \bibinfo {author} {\bibfnamefont
  {V.}~\bibnamefont {Durairaj}}, \bibinfo {author} {\bibfnamefont
  {G.}~\bibnamefont {Cao}}, \ and\ \bibinfo {author} {\bibfnamefont
  {E.}~\bibnamefont {Rotenberg}},\ }\href {\doibase
  10.1103/PhysRevLett.101.076402} {\bibfield  {journal} {\bibinfo  {journal}
  {Phys. Rev. Lett.}\ }\textbf {\bibinfo {volume} {101}},\ \bibinfo {pages}
  {076402} (\bibinfo {year} {2008})}\BibitemShut {NoStop}%
\bibitem [{\citenamefont {Witczak-Krempa}\ \emph {et~al.}(2014)\citenamefont
  {Witczak-Krempa}, \citenamefont {Chen}, \citenamefont {Kim},\ and\
  \citenamefont {Balents}}]{Witczak-Krempa2014}%
  \BibitemOpen
  \bibfield  {author} {\bibinfo {author} {\bibfnamefont {W.}~\bibnamefont
  {Witczak-Krempa}}, \bibinfo {author} {\bibfnamefont {G.}~\bibnamefont
  {Chen}}, \bibinfo {author} {\bibfnamefont {Y.~B.}\ \bibnamefont {Kim}}, \
  and\ \bibinfo {author} {\bibfnamefont {L.}~\bibnamefont {Balents}},\ }\href
  {\doibase 10.1146/annurev-conmatphys-020911-125138} {\bibfield  {journal}
  {\bibinfo  {journal} {Annual Review of Condensed Matter Physics}\ }\textbf
  {\bibinfo {volume} {5}},\ \bibinfo {pages} {57} (\bibinfo {year}
  {2014})}\BibitemShut {NoStop}%
\bibitem [{\citenamefont {Zhai}(2015)}]{Zhai2015}%
  \BibitemOpen
  \bibfield  {author} {\bibinfo {author} {\bibfnamefont {H.}~\bibnamefont
  {Zhai}},\ }\href@noop {} {\bibfield  {journal} {\bibinfo  {journal} {Reports
  on Progress in Physics}\ }\textbf {\bibinfo {volume} {78}},\ \bibinfo {pages}
  {026001} (\bibinfo {year} {2015})}\BibitemShut {NoStop}%
\bibitem [{\citenamefont {Dalibard}\ \emph {et~al.}(2011)\citenamefont
  {Dalibard}, \citenamefont {Gerbier}, \citenamefont
  {Juzeli\ifmmode~\bar{u}\else \={u}\fi{}nas},\ and\ \citenamefont
  {\"Ohberg}}]{DalibardRMP}%
  \BibitemOpen
  \bibfield  {author} {\bibinfo {author} {\bibfnamefont {J.}~\bibnamefont
  {Dalibard}}, \bibinfo {author} {\bibfnamefont {F.}~\bibnamefont {Gerbier}},
  \bibinfo {author} {\bibfnamefont {G.}~\bibnamefont
  {Juzeli\ifmmode~\bar{u}\else \={u}\fi{}nas}}, \ and\ \bibinfo {author}
  {\bibfnamefont {P.}~\bibnamefont {\"Ohberg}},\ }\href {\doibase
  10.1103/RevModPhys.83.1523} {\bibfield  {journal} {\bibinfo  {journal} {Rev.
  Mod. Phys.}\ }\textbf {\bibinfo {volume} {83}},\ \bibinfo {pages} {1523}
  (\bibinfo {year} {2011})}\BibitemShut {NoStop}%
\bibitem [{\citenamefont {Bauer}\ and\ \citenamefont {Sigrist}(2012)}]{NCSC}%
  \BibitemOpen
  \bibinfo {editor} {\bibfnamefont {E.}~\bibnamefont {Bauer}}\ and\ \bibinfo
  {editor} {\bibfnamefont {M.}~\bibnamefont {Sigrist}},\ eds.,\ \href@noop {}
  {\emph {\bibinfo {title} {Non-Centrosymmetric Superconductors: Introduction
  and Overview}}},\ Lecture Notes in Physics\ (\bibinfo  {publisher}
  {Springer},\ \bibinfo {year} {2012})\BibitemShut {NoStop}%
\bibitem [{\citenamefont {{Smidman}}\ \emph {et~al.}(2016)\citenamefont
  {{Smidman}}, \citenamefont {{Salamon}}, \citenamefont {{Yuan}},\ and\
  \citenamefont {{Agterberg}}}]{Smidman2016}%
  \BibitemOpen
  \bibfield  {author} {\bibinfo {author} {\bibfnamefont {M.}~\bibnamefont
  {{Smidman}}}, \bibinfo {author} {\bibfnamefont {M.~B.}\ \bibnamefont
  {{Salamon}}}, \bibinfo {author} {\bibfnamefont {H.~Q.}\ \bibnamefont
  {{Yuan}}}, \ and\ \bibinfo {author} {\bibfnamefont {D.~F.}\ \bibnamefont
  {{Agterberg}}},\ }\href@noop {} {\bibfield  {journal} {\bibinfo  {journal}
  {ArXiv e-prints}\ } (\bibinfo {year} {2016})},\ \Eprint
  {http://arxiv.org/abs/1609.05953} {arXiv:1609.05953 [cond-mat.supr-con]}
  \BibitemShut {NoStop}%
\bibitem [{\citenamefont {Caviglia}\ \emph {et~al.}(2010)\citenamefont
  {Caviglia}, \citenamefont {Gabay}, \citenamefont {Gariglio}, \citenamefont
  {Reyren}, \citenamefont {Cancellieri},\ and\ \citenamefont
  {Triscone}}]{Caviglia2010}%
  \BibitemOpen
  \bibfield  {author} {\bibinfo {author} {\bibfnamefont {A.~D.}\ \bibnamefont
  {Caviglia}}, \bibinfo {author} {\bibfnamefont {M.}~\bibnamefont {Gabay}},
  \bibinfo {author} {\bibfnamefont {S.}~\bibnamefont {Gariglio}}, \bibinfo
  {author} {\bibfnamefont {N.}~\bibnamefont {Reyren}}, \bibinfo {author}
  {\bibfnamefont {C.}~\bibnamefont {Cancellieri}}, \ and\ \bibinfo {author}
  {\bibfnamefont {J.-M.}\ \bibnamefont {Triscone}},\ }\href {\doibase
  10.1103/PhysRevLett.104.126803} {\bibfield  {journal} {\bibinfo  {journal}
  {Phys. Rev. Lett.}\ }\textbf {\bibinfo {volume} {104}},\ \bibinfo {pages}
  {126803} (\bibinfo {year} {2010})}\BibitemShut {NoStop}%
\bibitem [{\citenamefont {Berg}\ \emph {et~al.}(2012)\citenamefont {Berg},
  \citenamefont {Rudner},\ and\ \citenamefont {Kivelson}}]{Berg2012}%
  \BibitemOpen
  \bibfield  {author} {\bibinfo {author} {\bibfnamefont {E.}~\bibnamefont
  {Berg}}, \bibinfo {author} {\bibfnamefont {M.~S.}\ \bibnamefont {Rudner}}, \
  and\ \bibinfo {author} {\bibfnamefont {S.~A.}\ \bibnamefont {Kivelson}},\
  }\href {\doibase 10.1103/PhysRevB.85.035116} {\bibfield  {journal} {\bibinfo
  {journal} {Phys. Rev. B}\ }\textbf {\bibinfo {volume} {85}},\ \bibinfo
  {pages} {035116} (\bibinfo {year} {2012})}\BibitemShut {NoStop}%
\bibitem [{\citenamefont {Vafek}\ and\ \citenamefont {Wang}(2011)}]{VafekWang}%
  \BibitemOpen
  \bibfield  {author} {\bibinfo {author} {\bibfnamefont {O.}~\bibnamefont
  {Vafek}}\ and\ \bibinfo {author} {\bibfnamefont {L.}~\bibnamefont {Wang}},\
  }\href {\doibase 10.1103/PhysRevB.84.172501} {\bibfield  {journal} {\bibinfo
  {journal} {Phys. Rev. B}\ }\textbf {\bibinfo {volume} {84}},\ \bibinfo
  {pages} {172501} (\bibinfo {year} {2011})}\BibitemShut {NoStop}%
\bibitem [{\citenamefont {Wang}\ and\ \citenamefont {Vafek}(2014)}]{WangVafek}%
  \BibitemOpen
  \bibfield  {author} {\bibinfo {author} {\bibfnamefont {L.}~\bibnamefont
  {Wang}}\ and\ \bibinfo {author} {\bibfnamefont {O.}~\bibnamefont {Vafek}},\
  }\href {\doibase http://dx.doi.org/10.1016/j.physc.2013.10.007} {\bibfield
  {journal} {\bibinfo  {journal} {Physica C: Superconductivity and its
  Applications}\ }\textbf {\bibinfo {volume} {497}},\ \bibinfo {pages} {6 }
  (\bibinfo {year} {2014})}\BibitemShut {NoStop}%
\bibitem [{\citenamefont {Cappelluti}\ \emph {et~al.}(2007)\citenamefont
  {Cappelluti}, \citenamefont {Grimaldi},\ and\ \citenamefont
  {Marsiglio}}]{Cappelluti2007}%
  \BibitemOpen
  \bibfield  {author} {\bibinfo {author} {\bibfnamefont {E.}~\bibnamefont
  {Cappelluti}}, \bibinfo {author} {\bibfnamefont {C.}~\bibnamefont
  {Grimaldi}}, \ and\ \bibinfo {author} {\bibfnamefont {F.}~\bibnamefont
  {Marsiglio}},\ }\href {\doibase 10.1103/PhysRevLett.98.167002} {\bibfield
  {journal} {\bibinfo  {journal} {Phys. Rev. Lett.}\ }\textbf {\bibinfo
  {volume} {98}},\ \bibinfo {pages} {167002} (\bibinfo {year}
  {2007})}\BibitemShut {NoStop}%
\bibitem [{\citenamefont {Shankar}(1994)}]{Shankar}%
  \BibitemOpen
  \bibfield  {author} {\bibinfo {author} {\bibfnamefont {R.}~\bibnamefont
  {Shankar}},\ }\href {\doibase 10.1103/RevModPhys.66.129} {\bibfield
  {journal} {\bibinfo  {journal} {Rev. Mod. Phys.}\ }\textbf {\bibinfo {volume}
  {66}},\ \bibinfo {pages} {129} (\bibinfo {year} {1994})}\BibitemShut
  {NoStop}%
\bibitem [{\citenamefont {Alicea}(2012)}]{Alicea}%
  \BibitemOpen
  \bibfield  {author} {\bibinfo {author} {\bibfnamefont {J.}~\bibnamefont
  {Alicea}},\ }\href@noop {} {\bibfield  {journal} {\bibinfo  {journal}
  {Reports on Progress in Physics}\ }\textbf {\bibinfo {volume} {75}},\
  \bibinfo {pages} {076501} (\bibinfo {year} {2012})}\BibitemShut {NoStop}%
\bibitem [{\citenamefont {Kopnin}\ and\ \citenamefont
  {Salomaa}(1991)}]{KopninSalomaa}%
  \BibitemOpen
  \bibfield  {author} {\bibinfo {author} {\bibfnamefont {N.~B.}\ \bibnamefont
  {Kopnin}}\ and\ \bibinfo {author} {\bibfnamefont {M.~M.}\ \bibnamefont
  {Salomaa}},\ }\href {\doibase 10.1103/PhysRevB.44.9667} {\bibfield  {journal}
  {\bibinfo  {journal} {Phys. Rev. B}\ }\textbf {\bibinfo {volume} {44}},\
  \bibinfo {pages} {9667} (\bibinfo {year} {1991})}\BibitemShut {NoStop}%
\bibitem [{\citenamefont {Read}\ and\ \citenamefont {Green}(2000)}]{ReadGreen}%
  \BibitemOpen
  \bibfield  {author} {\bibinfo {author} {\bibfnamefont {N.}~\bibnamefont
  {Read}}\ and\ \bibinfo {author} {\bibfnamefont {D.}~\bibnamefont {Green}},\
  }\href {\doibase 10.1103/PhysRevB.61.10267} {\bibfield  {journal} {\bibinfo
  {journal} {Phys. Rev. B}\ }\textbf {\bibinfo {volume} {61}},\ \bibinfo
  {pages} {10267} (\bibinfo {year} {2000})}\BibitemShut {NoStop}%
\bibitem [{\citenamefont {Ivanov}(2001)}]{Ivanov2001}%
  \BibitemOpen
  \bibfield  {author} {\bibinfo {author} {\bibfnamefont {D.~A.}\ \bibnamefont
  {Ivanov}},\ }\href {\doibase 10.1103/PhysRevLett.86.268} {\bibfield
  {journal} {\bibinfo  {journal} {Phys. Rev. Lett.}\ }\textbf {\bibinfo
  {volume} {86}},\ \bibinfo {pages} {268} (\bibinfo {year} {2001})}\BibitemShut
  {NoStop}%
\bibitem [{\citenamefont {Tewari}\ \emph {et~al.}(2007)\citenamefont {Tewari},
  \citenamefont {Das~Sarma},\ and\ \citenamefont {Lee}}]{Tewari2007}%
  \BibitemOpen
  \bibfield  {author} {\bibinfo {author} {\bibfnamefont {S.}~\bibnamefont
  {Tewari}}, \bibinfo {author} {\bibfnamefont {S.}~\bibnamefont {Das~Sarma}}, \
  and\ \bibinfo {author} {\bibfnamefont {D.-H.}\ \bibnamefont {Lee}},\ }\href
  {\doibase 10.1103/PhysRevLett.99.037001} {\bibfield  {journal} {\bibinfo
  {journal} {Phys. Rev. Lett.}\ }\textbf {\bibinfo {volume} {99}},\ \bibinfo
  {pages} {037001} (\bibinfo {year} {2007})}\BibitemShut {NoStop}%
\bibitem [{\citenamefont {Fu}\ and\ \citenamefont {Kane}(2008)}]{FuKane2008}%
  \BibitemOpen
  \bibfield  {author} {\bibinfo {author} {\bibfnamefont {L.}~\bibnamefont
  {Fu}}\ and\ \bibinfo {author} {\bibfnamefont {C.~L.}\ \bibnamefont {Kane}},\
  }\href {\doibase 10.1103/PhysRevLett.100.096407} {\bibfield  {journal}
  {\bibinfo  {journal} {Phys. Rev. Lett.}\ }\textbf {\bibinfo {volume} {100}},\
  \bibinfo {pages} {096407} (\bibinfo {year} {2008})}\BibitemShut {NoStop}%
\bibitem [{\citenamefont {Sau}\ \emph {et~al.}(2010)\citenamefont {Sau},
  \citenamefont {Lutchyn}, \citenamefont {Tewari},\ and\ \citenamefont
  {Das~Sarma}}]{Sau2010PRL}%
  \BibitemOpen
  \bibfield  {author} {\bibinfo {author} {\bibfnamefont {J.~D.}\ \bibnamefont
  {Sau}}, \bibinfo {author} {\bibfnamefont {R.~M.}\ \bibnamefont {Lutchyn}},
  \bibinfo {author} {\bibfnamefont {S.}~\bibnamefont {Tewari}}, \ and\ \bibinfo
  {author} {\bibfnamefont {S.}~\bibnamefont {Das~Sarma}},\ }\href {\doibase
  10.1103/PhysRevLett.104.040502} {\bibfield  {journal} {\bibinfo  {journal}
  {Phys. Rev. Lett.}\ }\textbf {\bibinfo {volume} {104}},\ \bibinfo {pages}
  {040502} (\bibinfo {year} {2010})}\BibitemShut {NoStop}%
\bibitem [{\citenamefont {Mao}\ \emph {et~al.}(2011)\citenamefont {Mao},
  \citenamefont {Shi}, \citenamefont {Niu},\ and\ \citenamefont
  {Zhang}}]{Mao2011}%
  \BibitemOpen
  \bibfield  {author} {\bibinfo {author} {\bibfnamefont {L.}~\bibnamefont
  {Mao}}, \bibinfo {author} {\bibfnamefont {J.}~\bibnamefont {Shi}}, \bibinfo
  {author} {\bibfnamefont {Q.}~\bibnamefont {Niu}}, \ and\ \bibinfo {author}
  {\bibfnamefont {C.}~\bibnamefont {Zhang}},\ }\href {\doibase
  10.1103/PhysRevLett.106.157003} {\bibfield  {journal} {\bibinfo  {journal}
  {Phys. Rev. Lett.}\ }\textbf {\bibinfo {volume} {106}},\ \bibinfo {pages}
  {157003} (\bibinfo {year} {2011})}\BibitemShut {NoStop}%
\bibitem [{\citenamefont {Roy}(2010)}]{Roy2010PRL}%
  \BibitemOpen
  \bibfield  {author} {\bibinfo {author} {\bibfnamefont {R.}~\bibnamefont
  {Roy}},\ }\href {\doibase 10.1103/PhysRevLett.105.186401} {\bibfield
  {journal} {\bibinfo  {journal} {Phys. Rev. Lett.}\ }\textbf {\bibinfo
  {volume} {105}},\ \bibinfo {pages} {186401} (\bibinfo {year}
  {2010})}\BibitemShut {NoStop}%
\bibitem [{\citenamefont {Farrell}\ and\ \citenamefont
  {Pereg-Barnea}(2014)}]{Farrell}%
  \BibitemOpen
  \bibfield  {author} {\bibinfo {author} {\bibfnamefont {A.}~\bibnamefont
  {Farrell}}\ and\ \bibinfo {author} {\bibfnamefont {T.}~\bibnamefont
  {Pereg-Barnea}},\ }\href@noop {} {\bibfield  {journal} {\bibinfo  {journal}
  {Phys. Rev. B}\ }\textbf {\bibinfo {volume} {90}},\ \bibinfo {pages} {144518}
  (\bibinfo {year} {2014})}\BibitemShut {NoStop}%
\bibitem [{\citenamefont {Bogoliubov}\ \emph {et~al.}(1959)\citenamefont
  {Bogoliubov}, \citenamefont {Tolmachev},\ and\ \citenamefont
  {Shirkov}}]{Bogoliubov}%
  \BibitemOpen
  \bibfield  {author} {\bibinfo {author} {\bibfnamefont {N.~N.}\ \bibnamefont
  {Bogoliubov}}, \bibinfo {author} {\bibfnamefont {V.~V.}\ \bibnamefont
  {Tolmachev}}, \ and\ \bibinfo {author} {\bibfnamefont {D.~V.}\ \bibnamefont
  {Shirkov}},\ }\href@noop {} {\emph {\bibinfo {title} {A new method in the
  theory of superconductivity}}}\ (\bibinfo  {publisher} {Consultants Bureau},\
  \bibinfo {address} {New York},\ \bibinfo {year} {1959})\BibitemShut {NoStop}%
\bibitem [{\citenamefont {Anderson}(1958{\natexlab{a}})}]{Anderson1958a}%
  \BibitemOpen
  \bibfield  {author} {\bibinfo {author} {\bibfnamefont {P.~W.}\ \bibnamefont
  {Anderson}},\ }\href {\doibase 10.1103/PhysRev.110.827} {\bibfield  {journal}
  {\bibinfo  {journal} {Phys. Rev.}\ }\textbf {\bibinfo {volume} {110}},\
  \bibinfo {pages} {827} (\bibinfo {year} {1958}{\natexlab{a}})}\BibitemShut
  {NoStop}%
\bibitem [{\citenamefont {Anderson}(1958{\natexlab{b}})}]{Anderson1958b}%
  \BibitemOpen
  \bibfield  {author} {\bibinfo {author} {\bibfnamefont {P.~W.}\ \bibnamefont
  {Anderson}},\ }\href {\doibase 10.1103/PhysRev.112.1900} {\bibfield
  {journal} {\bibinfo  {journal} {Phys. Rev.}\ }\textbf {\bibinfo {volume}
  {112}},\ \bibinfo {pages} {1900} (\bibinfo {year}
  {1958}{\natexlab{b}})}\BibitemShut {NoStop}%
\bibitem [{\citenamefont {Leggett}(1966)}]{Leggett1966}%
  \BibitemOpen
  \bibfield  {author} {\bibinfo {author} {\bibfnamefont {A.~J.}\ \bibnamefont
  {Leggett}},\ }\href {\doibase 10.1143/PTP.36.901} {\bibfield  {journal}
  {\bibinfo  {journal} {Progress of Theoretical Physics}\ }\textbf {\bibinfo
  {volume} {36}},\ \bibinfo {pages} {901} (\bibinfo {year} {1966})}\BibitemShut
  {NoStop}%
\bibitem [{\citenamefont {Bittner}\ \emph {et~al.}(2015)\citenamefont
  {Bittner}, \citenamefont {Einzel}, \citenamefont {Klam},\ and\ \citenamefont
  {Manske}}]{Bittner2015}%
  \BibitemOpen
  \bibfield  {author} {\bibinfo {author} {\bibfnamefont {N.}~\bibnamefont
  {Bittner}}, \bibinfo {author} {\bibfnamefont {D.}~\bibnamefont {Einzel}},
  \bibinfo {author} {\bibfnamefont {L.}~\bibnamefont {Klam}}, \ and\ \bibinfo
  {author} {\bibfnamefont {D.}~\bibnamefont {Manske}},\ }\href@noop {}
  {\bibfield  {journal} {\bibinfo  {journal} {Phys. Rev. Lett.}\ }\textbf
  {\bibinfo {volume} {115}},\ \bibinfo {pages} {227002} (\bibinfo {year}
  {2015})}\BibitemShut {NoStop}%
\bibitem [{\citenamefont {Kohn}\ and\ \citenamefont
  {Luttinger}(1965)}]{Kohn1965}%
  \BibitemOpen
  \bibfield  {author} {\bibinfo {author} {\bibfnamefont {W.}~\bibnamefont
  {Kohn}}\ and\ \bibinfo {author} {\bibfnamefont {J.~M.}\ \bibnamefont
  {Luttinger}},\ }\href@noop {} {\bibfield  {journal} {\bibinfo  {journal}
  {Phys. Rev. Lett.}\ }\textbf {\bibinfo {volume} {15}},\ \bibinfo {pages}
  {524} (\bibinfo {year} {1965})}\BibitemShut {NoStop}%
\bibitem [{\citenamefont {{Sharapov, S. G.}}\ \emph {et~al.}(2002)\citenamefont
  {{Sharapov, S. G.}}, \citenamefont {{Gusynin, V. P.}},\ and\ \citenamefont
  {{Beck, H.}}}]{Sharapov2002}%
  \BibitemOpen
  \bibfield  {author} {\bibinfo {author} {\bibnamefont {{Sharapov, S. G.}}},
  \bibinfo {author} {\bibnamefont {{Gusynin, V. P.}}}, \ and\ \bibinfo {author}
  {\bibnamefont {{Beck, H.}}},\ }\href@noop {} {\bibfield  {journal} {\bibinfo
  {journal} {Eur. Phys. J. B}\ }\textbf {\bibinfo {volume} {30}},\ \bibinfo
  {pages} {45} (\bibinfo {year} {2002})}\BibitemShut {NoStop}%
\end{thebibliography}%

\begin{widetext}
\section{supplemental material}
\renewcommand{\theequation}{S\arabic{equation}}
\setcounter{equation}{0}
\subsection{Appendix: Collective modes in the superconducting state}
In the main text, we have found that the superconducting state of a 2D repulsive Rashba gas at low density breaks time reversal symmetry, and the Cooper pair has a total angular momentum $j_z=2$. In the superconducting state, we replace the full interaction with its projection onto $j_z=2$ channel, and get the following Hamiltonian:
\begin{eqnarray}
H&=&H_{kin}+H_{int},\\
H_{kin}&=&\sum_{\bk\lambda}(E_{\bk\lambda}-\mu)a_{\bk\lambda}^\dagger a_{\bk\lambda} =\sum_{\bk\lambda}\xi_{\bk\lambda}a_{\bk\lambda}^\dagger a_{\bk\lambda},\\
H_{int}&=&\sum_{\bk\bk'}\sum_{\mu\lambda}g_{\mu\lambda}e^{i(\theta_{\bk}-\theta_{\bk'})}a_{\bk\mu}^\dagger a_{-\bk\mu}^\dagger a_{-\bk'\lambda}a_{\bk'\lambda},
\end{eqnarray}
where $g_{\mu\lambda}\equiv\frac{u^2m}{32}V_{\mu\lambda}^{(2)}$. We apply path integral formalism to study the collective modes of this system, following Ref.\cite{Sharapov2002}. The partition function is
\begin{eqnarray}
Z=\int D\psi^\dagger_{\lambda}D\psi_{\lambda}e^{-S_0-S_{int}},
\end{eqnarray}
where
\begin{eqnarray}
S_0&=&\int_0^\be d\tau\left[\sum_{\bk\lambda}\psi^\dagger_{\lambda}(\tau,\bk)(\partial_\tau+\xi_{\bk\lambda}) \psi_{\lambda}(\tau,\bk)\right],\\
S_{int}&=&\int_0^\be d\tau\sum_{\bk\bk'\mu\lambda}g_{\mu\lambda}e^{i(\theta_\bk-\theta_{\bk'})}\psi^\dagger_{\mu}(\tau,\bk) \psi^\dagger_{\mu}(\tau,-\bk)\psi_{\lambda}(\tau,-\bk')\psi_{\lambda}(\tau,\bk').
\end{eqnarray}
Now we go back to real space where the operator $\psi_\lambda$ can be separated into modulus and phase variables, so that the phase fluctuations are convenient to study. Since we were only considering the Cooper channel and the total incoming and outgoing momenta are 0, when Fourier transforming to real space, we need to allow a finite total momentum $\bq$, i.e.
\begin{eqnarray}
S_{int}&=&\int_0^\be d\tau\sum_{\bk\bk'\bq\mu\lambda}g_{\mu\lambda}e^{i(\theta_\bk-\theta_{\bk'})}\psi^\dagger_{\mu}(\tau,\bk) \psi^\dagger_{\mu}(\tau,-\bk+\bq)\psi_{\lambda}(\tau,-\bk'+\bq)\psi_{\lambda}(\tau,\bk').
\end{eqnarray}
The phase $e^{i\theta_\bk}=\frac{k_x+ik_y}{k_{F\mu}}$ in real space transforms to $-i\frac{\partial_x+i\partial_y}{k_{F\mu}}$ and $\xi_{k\lambda}$ transforms to
\begin{eqnarray}
\xi_\lambda(-i\nabla)=-\frac{\nabla^2}{2m}-\lambda\al_R|-i\nabla|-\mu,
\end{eqnarray}
where in the second term, the symbol ``$|\ |$" means the amplitude of a vector, but leaves its (complex) coefficient untouched.
Then
\begin{eqnarray}
S_0&=&\int_0^\be d\tau\int d\br\sum_{\lambda}\psi^\dagger_{\lambda}(\tau,\br)(\partial_\tau+\xi_\lambda(-i\nabla)) \psi_{\lambda}(\tau,\br),\\
S_{int}&=&\int_0^\be d\tau\int d\br\sum_{\mu\lambda}\frac{g_{\mu\lambda}}{k_{F\mu}k_{F\lambda}}\psi^\dagger_{\mu}(\tau,\br) [-i(\partial_x+i\partial_y)\psi^\dagger_{\mu}(\tau,\br)] \psi_{\lambda}(\tau,\br)[-i(\partial_x-i\partial_y)\psi_{\lambda}(\tau,\br)].
\end{eqnarray}
Introducing Nambu spinors
\begin{eqnarray}
\Psi_\lambda(\tau,\br)=\left(\begin{array}{c}\psi_\lambda(\tau,\br)\\\psi_\lambda^\dagger(\tau,\br)\end{array}\right),
\end{eqnarray}
the action becomes
\begin{eqnarray}
S_0&=&\frac{1}{2}\int_0^\be d\tau\int d\br\sum_{\lambda}\Psi^\dagger_{\lambda}(\tau,\br)(\partial_\tau+\tau_3\xi_\lambda(-i\nabla)) \Psi_{\lambda}(\tau,\br),\\
S_{int}&=&\int_0^\be d\tau\int d\br\sum_{\mu\lambda}g_{\mu\lambda}\Psi_{\mu}^\dagger(\tau,\br)\tau_+[-i(\partial_x+i\partial_y)\Psi_{\mu}(\tau,\br)]  \Psi^\dagger_{\lambda}(\tau,\br)\tau_-[-i(\partial_x-i\partial_y)\Psi_{\lambda}(\tau,\br)].
\end{eqnarray}
Here $\tau_i$ are Pauli matrices and $\tau_\pm=\frac{1}{2}(\tau_1\pm i\tau_2)$, and $1/k_{F\mu}k_{F\lambda}$ is absorbed into $g_{\mu\lambda}$. We multiply the ``fat identity" $\int D\Phi^*D\Phi e^{-S_\Phi}$ to the partition function $Z$, where
\begin{eqnarray}
S_\Phi&=&-\int_0^\be d\tau\int d\br\sum_{\mu\lambda}g_{\mu\lambda}\Phi_\mu^*(\tau,\br)\Phi_\lambda(\tau,\br)\\
&=&-\int_0^\be d\tau\int d\br\sum_{\mu\lambda}g_{\mu\lambda}\nonumber\\
&&\left\{\Phi_\mu^*(\tau,\br)+\Psi_{\mu}^\dagger(\tau,\br)\tau_+[-i(\partial_x+i\partial_y)\Psi_{\mu}(\tau,\br)]\right\} \left\{\Phi_\lambda(\tau,\br)+\Psi^\dagger_{\lambda}(\tau,\br)\tau_-[-i(\partial_x-i\partial_y)\Psi_{\lambda}(\tau,\br)]\right\},
\end{eqnarray}
then we have a new action describing the pairing field $\Phi$,
\begin{eqnarray}
S_{pair}&=&S_{int}+S_\Phi\\
&=&-\int_0^\be d\tau\int d\br\sum_{\mu\lambda} \left(g_{\mu\lambda}\Phi_\mu^*\Phi_\lambda+g_{\mu\lambda}\Phi_\mu^*\Psi_\lambda^\dagger\tau_-[-i(\partial_x-i\partial_y)\Psi_{\lambda}] +g_{\mu\lambda}\Phi_\lambda\Psi_{\mu}^\dagger\tau_+[-i(\partial_x+i\partial_y)\Psi_{\mu}]\right).
\end{eqnarray}
Note that the bosonic field $\Phi$ in one band now couples not only to the fermionic field $\Psi$ in the same band, but also to that in the other band. We can make a linear transformation to avoid this. From now on, we change our notation: we use 1 and 2 to label the outer Fermi surface and inner Fermi surface, respectively, instead of $+1$ and $-1$. Let the coefficient of $\Psi_{\mu}^\dagger\tau_+[-i(\partial_x+i\partial_y)\Psi_{\mu}]$ be $\Delta_\mu=\sum_\lambda g_{\mu\lambda}\Phi_\lambda$, or in matrix form,
\begin{eqnarray}
\left(\begin{array}{c}\Delta_1\\\Delta_2\end{array}\right)=\left(\begin{array}{cc}g_{11}&g_{12}\\g_{12}&g_{22}\end{array}\right) \left(\begin{array}{c}\Phi_1\\\Phi_2\end{array}\right).
\end{eqnarray}
The inverse transformation is
\begin{eqnarray}
\left(\begin{array}{c}\Phi_1\\\Phi_2\end{array}\right)=\frac{1}{g_{11}g_{22}-g_{12}^2} \left(\begin{array}{cc}g_{22}&-g_{12}\\-g_{12}&g_{11}\end{array}\right) \left(\begin{array}{c}\Delta_1\\\Delta_2\end{array}\right),
\end{eqnarray}
then the first term in $S_{pair}$ becomes
\begin{eqnarray}
\sum_{\mu\lambda}g_{\mu\lambda}\Phi_\mu^*\Phi_\lambda=(\Phi_1^*,\Phi_2^*)\left(\begin{array}{c}\Delta_1\\\Delta_2\end{array}\right) &=&\frac{g_{22}}{g_{11}g_{22}-g_{12}^2}|\Delta_1|^2+\frac{g_{11}}{g_{11}g_{22}-g_{12}^2}|\Delta_2|^2 -\frac{g_{12}}{g_{11}g_{22}-g_{12}^2}(\Delta_1^*\Delta_2+\Delta_2^*\Delta_1)\\
&=&\frac{1}{g'_{11}}|\Delta_1|^2+\frac{1}{g'_{22}}|\Delta_2|^2+\frac{1}{g_{12}'}(\Delta_1^*\Delta_2+\Delta_2^*\Delta_1),
\end{eqnarray}
where
\begin{eqnarray}
g_{11}'=g_{11}(1-\frac{g_{12}^2}{g_{11}g_{22}}),g_{22}'=g_{22}(1-\frac{g_{12}^2}{g_{11}g_{22}}), g_{12}'=g_{12}(1-\frac{g_{11}g_{22}}{g_{12}^2}).
\end{eqnarray}
Therefore, the pairing action can be written as
\begin{eqnarray}
S_{pair}&=&-\int_0^\be d\tau\int d\br\left\{\sum_{\mu\lambda} \frac{1}{g'_{\mu\lambda}}\Delta_\mu^*\Delta_\lambda+\sum_\lambda\left[\Delta_\lambda^* \Psi_\lambda^\dagger\tau_-[-i(\partial_x-i\partial_y)\Psi_{\lambda}] +\Delta_\lambda\Psi_{\lambda}^\dagger\tau_+[-i(\partial_x+i\partial_y)\Psi_{\lambda}]\right]\right\},
\end{eqnarray}
and the partition function becomes $Z=\int D\Psi^\dagger D\Psi D\Delta^* D\Delta e^{-S_0-S_{pair}}$.

We introduce the modulus-phase variables as follows:
\begin{eqnarray}
\Delta_{\lambda}(\tau,\br)&=&|\Delta_\lambda(\tau,\br)|e^{i\theta_\lambda(\tau,\br)},\\
\Psi_\lambda(\tau,\br)&=&\left(\begin{array}{cc}e^{i\theta_\lambda(\tau,\br)/2}&0\\0&e^{-i\theta_\lambda(\tau,\br)/2}\end{array}\right) \Upsilon_\lambda(\tau,\br)=e^{\tau_3 \frac{i\theta_\lambda(\tau,\br)}{2}}\Upsilon_\lambda(\tau,\br),
\end{eqnarray}
and use the real field $\Delta$ to denote $|\Delta|$. Then
\begin{eqnarray}
Z=\int \Delta_\lambda D\Delta_\lambda D\theta_\lambda D\Upsilon_\lambda^\dagger D\Upsilon_\lambda e^{-S_0\{\Upsilon_\lambda,\Upsilon_\lambda^\dagger,\theta_\lambda\}-S_{pair}\{\Delta_\lambda,\theta_\lambda,\Upsilon_\lambda, \Upsilon_\lambda^\dagger\}}.
\end{eqnarray}
It is straight forward to show that the action in terms of modulus-phase variables is
\begin{eqnarray}
S_0&=&\frac{1}{2}\int_0^\be d\tau\int d\br\sum_\lambda\Upsilon_\lambda^\dagger\left\{\partial_\tau- \tau_3(\frac{\nabla^2}{2m}+\mu)+\tau_3[\frac{i\partial_\tau\theta_\lambda}{2}+\frac{(\nabla\theta_\lambda)^2}{8m}- \lambda\al_R|-i\nabla|]\right.\nonumber\\
&&\left.-[\frac{i\nabla\theta_\lambda\cdot\nabla}{2m}+\frac{i\nabla^2\theta_\lambda}{4m} +\frac{\lambda\al_R|\nabla\theta_\lambda|}{2}]\right\}\Upsilon_\lambda,\\
S_{pair}&=&-\int_0^\be d\tau\int d\br\left\{\frac{\Delta_1^2}{g_{11}'}+\frac{\Delta_2^2}{g_{22}'}+\frac{2}{g_{12}'} \Delta_1\Delta_2\cos(\theta_1-\theta_2)\right.\nonumber\\
&+&\left.\sum_\lambda\Upsilon_\lambda^\dagger\Delta_\lambda \left(\tau_-[-i(\partial_x-i\partial_y)+\frac{1}{2}(\partial_x\theta_\lambda-i\partial_y\theta_\lambda)] +\tau_+[-i(\partial_x+i\partial_y)+\frac{1}{2}(-\partial_x\theta_\lambda-i\partial_y\theta_\lambda)]\right)\Upsilon_\lambda\right\}.
\end{eqnarray}
Integrating out neutral fermions $\Upsilon_\lambda$, we obtain $Z=\int \Delta_\lambda D\Delta_\lambda D\theta_\lambda e^{-S_{eff}\{\Delta_\lambda,\theta_\lambda\}}$, with the effective action
\begin{eqnarray}
S_{eff}=-\int_0^\be d\tau d\br\left[\frac{\Delta_1^2}{g_{11}'}+\frac{\Delta_2^2}{g_{22}'}+\frac{2}{g_{12}'} \Delta_1\Delta_2\cos(\theta_1-\theta_2)\right]-\mbox{Tr}\ln G_1^{-1}-\mbox{Tr}\ln G_2^{-1},
\end{eqnarray}
where the Green's function is
\begin{eqnarray}
G_\lambda^{-1}=G_{\lambda}^{(0)-1}-\Sigma_\lambda(\partial\theta_\lambda),
\end{eqnarray}
with
\begin{eqnarray}
G_{\lambda}^{(0)-1}&=&\frac{1}{2}\left[\partial_\tau-\tau_3(\frac{\nabla^2}{2m}+\lambda\al_R|-i\nabla|+\mu)\right]-\Delta_\lambda \left(\tau_-[-i(\partial_x-i\partial_y)]+\tau_+[-i(\partial_x+i\partial_y)]\right),\\
\Sigma_\lambda(\partial\theta_\lambda)&=&-\frac{1}{2}\left\{\tau_3[\frac{i\partial_\tau\theta_\lambda}{2} +\frac{(\nabla\theta_\lambda)^2}{8m}]-[\frac{i\nabla\theta_\lambda\cdot\nabla}{2m}+\frac{i\nabla^2\theta_\lambda}{4m} +\frac{\lambda\al_R|\nabla\theta_\lambda|}{2}]\right.\nonumber\\
&&\left.-\Delta_\lambda[\tau_-(\partial_x\theta_\lambda-i\partial_y\theta_\lambda)+ \tau_+(-\partial_x\theta_\lambda-i\partial_y\theta_\lambda)]\right\}.
\end{eqnarray}

Now that the modulus and phase variables are separated, we can go to frequency-momentum space by the Fourier transform
\begin{eqnarray}
\Upsilon_\lambda(\tau,\br)&=&\frac{1}{\sqrt{\be}}\sum_{\omega_n,\bk}\Upsilon_\lambda(i\omega_n,\bk)e^{i\bk\cdot\br-i\omega_n\tau},\\
\theta_\lambda(\tau,\br)&=&\frac{1}{\sqrt{\be}}\sum_{\Omega_n,\bK}\theta_\lambda(i\Omega_n,\bK)e^{i\bK\cdot\br-i\Omega_n\tau},
\end{eqnarray}
where $\omega_n=(2n+1)\frac{\pi}{\be}$ and $\Omega_n=2n\frac{\pi}{\be}$. We fix $\Delta_\lambda$'s to the value derived by the saddle point approximation, since we only consider the phase fluctuations. Then the terms $\frac{\Delta_\lambda^2}{g_{\lambda\lambda}'}$ can be dropped, and the Josephson term is expanded to second order of $\theta_\lambda$:
\begin{eqnarray}
-\int d\tau d\br\frac{2}{g_{12}'}\Delta_1\Delta_2\cos(\theta_1-\theta_2)\to\frac{\Delta_1\Delta_2}{g_{12}'} \sum_{\Omega_n,\bK}[\theta_1(i\Omega_n,\bK)-\theta_2(i\Omega_n,\bK)][\theta_1(-i\Omega_n,-\bK)-\theta_2(-i\Omega_n,-\bK)].
\end{eqnarray}
We write other terms of the action in two parts $S_0'+S_\theta$, with $S_\theta$ including the dynamics of $\theta_\lambda$ while $S_0'$ not; and $S_\theta$ has a first order term and a second order term:
\begin{eqnarray}
S_0'&=&\frac{1}{2}\sum_\lambda\sum_{\omega_n,\bk}\Upsilon_\lambda^\dagger(i\omega_n,\bk)[-i\omega_n+\tau_3\xi_{\bk\lambda} -2\Delta_\lambda(\tau_- e^{-i\theta_\bk}+\tau_+ e^{i\theta_\bk})]\Upsilon_\lambda(i\omega_n,\bk),\\
S_{\theta}^{(1)}&=&\frac{1}{2\sqrt{\be}}\sum_\lambda\sum_{\omega_l,\bk}\sum_{\Omega_n,\bK} \Upsilon_\lambda^\dagger(i\omega_l+i\Omega_n,\bk+\bK)\theta_\lambda(i\Omega_n,\bK)\left\{\frac{\tau_3\Omega_n}{2}+ \frac{i\bK\cdot\bk}{2m}+\frac{iK^2}{4m}+\frac{i\lambda\al_R K}{2}\right.\nonumber\\
&&\left.-i\Delta_\lambda[(K_x-iK_y)\tau_-+(K_x+iK_y)\tau_+]\right\} \Upsilon_\lambda(i\omega_l,\bk),\\
&=&\frac{1}{2\sqrt{\be}}\sum_\lambda\sum_{\omega_l,\bk}\sum_{\Omega_n,\bK} \Upsilon_\lambda^\dagger(i\omega_l+i\Omega_n,\bk+\bK)\theta(i\Omega_n,\bK)_\lambda\left[\frac{\tau_3\Omega_n}{2}+ \frac{i\bK\cdot\bk}{2m}+\frac{iK^2}{4m}+\frac{i\lambda\al_R K}{2}\right.\nonumber\\
&&\left.-i\Delta_\lambda(K_x\tau_1-K_y\tau_2)\right] \Upsilon_\lambda(i\omega_l,\bk),\\
S_{\theta}^{(2)}&=&\frac{1}{2\be}\sum_{\lambda}\sum_{\omega_l,\bk}\sum_{\Omega_n,\bK_1}\sum_{\Omega_m,\bK_2} \Upsilon_\lambda^\dagger(i\omega_l+i\Omega_n+i\Omega_m,\bk+\bK_1+\bK_2)\nonumber\\ &&\left[\tau_3\frac{\theta_\lambda(i\Omega_n,K_1)\theta_\lambda(i\Omega_m,K_2)}{8m}(-K_1K_2)\right]\Upsilon_\lambda(i\omega_l,\bk).
\end{eqnarray}

Treating $S_\theta$ as a perturbation, we have the cumulant expansion
\begin{eqnarray}
\int D\Upsilon_\lambda^\dagger D\Upsilon_\lambda e^{-S_0'-S_\theta}=\langle e^{-S_\theta}\rangle\equiv e^{-S_{e}}\approx e^{-\langle S_\theta\rangle+\frac{1}{2}(\langle S_\theta^2\rangle-\langle S_\theta\rangle^2)}.
\end{eqnarray}
where the average is with respect to $S_0'$ and $S_e$ is the effective action for $\theta_\lambda$. We will keep $S_{e}$ to the second order of $\theta_\lambda$. The common factor $\frac{1}{2}$ in $S_0'$ and $S_\theta$ can be omitted, since it gives a constant term in $S_e$. It is easy to see $\langle S_\theta^{(1)}\rangle=0$, while
\begin{eqnarray}
\langle S_\theta^{(2)}\rangle&=&\frac{1}{8m\be}\sum_{\lambda}\left[\sum_{\Omega_n,\bK} \theta_\lambda(i\Omega_n,\bK)K^2\theta_\lambda(-i\Omega_n,-\bK)\sum_{\omega_l,\bk}\langle \Upsilon_\lambda^\dagger(i\omega_l,\bk)\tau_3\Upsilon_\lambda(i\omega_l,\bk)\rangle\right]\\
&=&-\frac{1}{8m\be}\sum_{\lambda}\left[\sum_{\Omega_n,\bK} \theta_\lambda(i\Omega_n,\bK)K^2\theta_\lambda(-i\Omega_n,-\bK)\mbox{Tr}(G_\lambda^{(0)}\tau_3)\right].
\end{eqnarray}
The trace is
\begin{eqnarray}
-\frac{1}{\be}\mbox{Tr}(G_\lambda^{(0)}\tau_3)&=&\frac{1}{\be}\sum_{\omega_l,\bk}\mbox{tr}(G_\lambda^{(0)}(i\omega_l,\bk)\tau_3)\\
&=&-\frac{1}{\be}\sum_{\omega_l,\bk}\mbox{tr}[\frac{1}{-i\omega_n+\tau_3\xi_{\bk\lambda} -2\Delta_\lambda(\tau_- e^{-i\theta_\bk}+\tau_+ e^{i\theta_\bk})}\tau_3]\\
&=&-\frac{1}{\be}\sum_{\omega_l,\bk}\mbox{tr}[\frac{i\omega_n+\tau_3\xi_{\bk\lambda}-2\Delta_\lambda(\tau_- e^{-i\theta_\bk}+\tau_+ e^{i\theta_\bk})} {\omega_n^2+\xi^2_{\bk\lambda}+4\Delta_\lambda^2}\tau_3]\\
&=&-\frac{2}{\be}\sum_{\omega_l,\bk}\frac{\xi_{\bk\lambda}} {\omega_n^2+\xi^2_{\bk\lambda}+4\Delta_\lambda^2}\\
&=&\int\frac{d\bk}{(2\pi)^2}\left(1-\frac{\xi_{\bk\lambda}}{E_{\bk\lambda}}\tanh\frac{\be E_{\bk\lambda}}{2}\right)\\
&=&2n_\lambda,
\end{eqnarray}
where $n_\lambda$ at $T=0$ is the total electron number in band $\lambda$. Then
\begin{eqnarray}
\langle S_\theta^{(2)}\rangle&=&\frac{1}{8}\sum_{\lambda}\left[\sum_{\Omega_n,\bK} \theta_\lambda(i\Omega_n,\bK)\frac{2n_\lambda}{m}K^2\theta_\lambda(-i\Omega_n,-\bK)\right].
\end{eqnarray}
We keep up to quadratic terms of $\theta_\lambda$, so
\begin{eqnarray}
\frac{1}{2}\langle S_\theta^2\rangle&\approx&\frac{1}{2}\langle S_\theta^{(1)2}\rangle\\
&=&\frac{1}{2\be}\sum_\lambda\langle\sum_{\omega_l,\bk}\sum_{\Omega_n,\bK} \Upsilon_\lambda^\dagger(i\omega_l+i\Omega_n,\bk+\bK)\theta_\lambda(i\Omega_n,\bK)\left\{\frac{\tau_3\Omega_n}{2}+ \frac{i\bK\cdot\bk}{2m}+\frac{iK^2}{4m}+\frac{i\lambda\al_R K}{2}\right.\nonumber\\
&&\left.-i\Delta_\lambda(K_x\tau_1-K_y\tau_2)\right\} \Upsilon_\lambda(i\omega_l,\bk)\nonumber\\
&&\times\sum_{\omega_{l'},\bk'}\sum_{\Omega_{n'},\bK'} \Upsilon_\lambda^\dagger(i\omega_{l'}+i\Omega_{n'},\bk'+\bK')\theta_\lambda(i\Omega_{n'},\bK')\left\{\frac{\tau_3\Omega_{n'}}{2}+ \frac{i\bK'\cdot\bk'}{2m}+\frac{iK'^2}{4m}+\frac{i\lambda\al_R K'}{2}\right.\nonumber\\
&&\left.-i\Delta_\lambda(K'_x\tau_1-K'_y\tau_2)\right\} \Upsilon_\lambda(i\omega_{l'},\bk')\rangle\\
&=&\frac{1}{2\be}\sum_\lambda\sum_{\omega_l,\bk}\sum_{\Omega_n,\bK}\sum_{\omega_{l'},\bk'}\sum_{\Omega_{n'},\bK'} \theta_\lambda(i\Omega_n,\bK)\theta_\lambda(i\Omega_{n'},\bK')\nonumber\\
&&\langle \Upsilon_\lambda^\dagger(i\omega_l+i\Omega_n,\bk+\bK)\left\{\frac{\tau_3\Omega_n}{2}+ \frac{i\bK\cdot\bk}{2m}+\frac{iK^2}{4m}+\frac{i\lambda\al_R K}{2}-i\Delta_\lambda(K_x\tau_1-K_y\tau_2)\right\} \Upsilon_\lambda(i\omega_l,\bk)\nonumber\\
&&\Upsilon_\lambda^\dagger(i\omega_{l'}+i\Omega_{n'},\bk'+\bK')\left\{\frac{\tau_3\Omega_{n'}}{2}+ \frac{i\bK'\cdot\bk'}{2m}+\frac{iK'^2}{4m}+\frac{i\lambda\al_R K'}{2}-i\Delta_\lambda(K'_x\tau_1-K'_y\tau_2)\right\} \Upsilon_\lambda(i\omega_{l'},\bk')\rangle\\
&=&\frac{1}{2\be}\sum_\lambda\sum_{\omega_l,\bk}\sum_{\Omega_n,\bK}
\theta_\lambda(i\Omega_n,\bK)\theta_\lambda(-i\Omega_{n},-\bK)\nonumber\\
&&\langle \Upsilon_\lambda^\dagger(i\omega_l+i\Omega_n,\bk+\bK)\left\{\frac{\tau_3\Omega_n}{2}+ \frac{i\bK\cdot\bk}{2m}+\frac{iK^2}{4m}+\frac{i\lambda\al_R K}{2}-i\Delta_\lambda(K_x\tau_1-K_y\tau_2)\right\} \Upsilon_\lambda(i\omega_l,\bk)\nonumber\\
&&\Upsilon_\lambda^\dagger(i\omega_{l},\bk)\left\{-\frac{\tau_3\Omega_{n}}{2}- \frac{i\bK\cdot(\bk+\bK)}{2m}+\frac{iK^2}{4m}-\frac{i\lambda\al_R K}{2}+i\Delta_\lambda(K_x\tau_1-K_y\tau_2)\right\} \Upsilon_\lambda(i\omega_{l}+i\Omega_n,\bk+\bK)\rangle\\
&\approx&-\frac{1}{2}\sum_\lambda\sum_{\Omega_n,\bK}
\theta_\lambda(i\Omega_n,\bK)\theta_\lambda(-i\Omega_{n},-\bK)\nonumber\\
&&\frac{1}{\be}\sum_{\omega_l,\bk}\mbox{tr}\left\{G^{(0)}_\lambda(i\omega_l+i\Omega_n,\bk+\bK)\left[\frac{\tau_3\Omega_n}{2}+ \frac{i\bK\cdot\bk}{2m}+\frac{i\lambda\al_R K}{2}-i\Delta_\lambda(K_x\tau_1-K_y\tau_2)\right]\right.\nonumber\\
&&\left.G^{(0)}_\lambda(i\omega_{l},\bk)\left[-\frac{\tau_3\Omega_{n}}{2}- \frac{i\bK\cdot\bk}{2m}-\frac{i\lambda\al_R K}{2}+i\Delta_\lambda(K_x\tau_1-K_y\tau_2)\right]\right\}.
\end{eqnarray}
If $\al_R=0$, this corresponds to the case of two-band $p+ip$ superconductors,
\begin{eqnarray}
\frac{1}{2}\langle S_\theta^2\rangle&\approx&-\frac{1}{8}\sum_\lambda\sum_{\Omega_n,\bK}\theta_\lambda(i\Omega_n,\bK) M_\lambda^{-1}\theta_\lambda(-i\Omega_n,-\bK),
\end{eqnarray}
where
\begin{eqnarray}
M_\lambda^{-1}&=&-\Omega_n^2{}^\lambda\Pi_{33}(i\Omega_n,\bK)+K_\al K_\be{}^\lambda\Pi_{00}^{\al\be}(i\Omega_n,\bK)-i\Omega_nK_\al[^\lambda\Pi_{03}^\al(i\Omega_n,\bK) +{}^\lambda\Pi_{30}^\al(i\Omega_n,\bK)]\nonumber\\
&&+4\Delta_\lambda^2[K_x^2{}^\lambda\Pi_{11}(i\Omega_n,\bK) +K_y^2{}^\lambda\Pi_{22}(i\Omega_n,\bK)-K_xK_y{}^\lambda\Pi_{12}(i\Omega_n,\bK) -K_xK_y{}^\lambda\Pi_{21}(i\Omega_n,\bK)]\nonumber\\
&&+2i\Delta_\lambda\Omega_n K_x[^\lambda\Pi_{31}(i\Omega_n,\bK)+{}^\lambda\Pi_{13}(i\Omega_n,\bK)]-2i\Delta_\lambda\Omega_n K_y[^\lambda\Pi_{32}(i\Omega_n,\bK)+{}^\lambda\Pi_{23}(i\Omega_n,\bK)]\nonumber\\
&&-2\Delta_\lambda K_xK_\al[^\lambda\Pi_{01}^\al(i\Omega_n,\bK)+{}^\lambda\Pi_{10}^\al(i\Omega_n,\bK)] +2\Delta_\lambda K_yK_\al[^\lambda\Pi_{02}^\al(i\Omega_n,\bK)+{}^\lambda\Pi_{20}^\al(i\Omega_n,\bK)],
\end{eqnarray}
in which we have defined
\begin{eqnarray}
^\lambda\Pi_{ij}(i\Omega_n,\bK)&\equiv&\frac{1}{\be}\sum_{\omega_l,\bk}{}^\lambda\pi_{ij}(i\Omega_n,\bK;i\omega_l,\bk),\\
^\lambda\Pi_{ij}^{\al}(i\Omega_n,\bK)&\equiv&\frac{1}{\be}\sum_{\omega_l,\bk}{}^\lambda\pi_{ij}(i\Omega_n,\bK;i\omega_l,\bk) v_{F\lambda\al}(\bk),\\
^\lambda\Pi_{ij}^{\al\be}(i\Omega_n,\bK)&\equiv&\frac{1}{\be}\sum_{\omega_l,\bk}{}^\lambda\pi_{ij}(i\Omega_n,\bK;i\omega_l,\bk) v_{F\lambda\al}(\bk)v_{F\lambda\be}(\bk),
\end{eqnarray}
with $v_{F\lambda\al}(\bk)=\partial\xi_\lambda(\bk)/\partial k_\al|_{k=k_{F\lambda}}$, and
\begin{eqnarray}
{}^\lambda\pi_{ij}(i\Omega_n,\bK;i\omega_l,\bk)\equiv\mbox{tr}[G^{(0)}_\lambda(i\omega_l+i\Omega_n,\bk+\bK)\tau_i G^{(0)}_\lambda(i\omega_{l},\bk)\tau_j],\ \ \ (\tau_0\equiv I).
\end{eqnarray}
Now we need to evaluate the 16 correlation functions. Using
\begin{eqnarray}
\frac{1}{\be}\sum_{\omega_l}h(\omega_l)=\sum_{j}\mbox{Res}\ h(-iz)n_F(z)\vert_{z=z_j},
\end{eqnarray}
it is straightforward to do the Matsubara sum,
\begin{eqnarray}
&&\frac{1}{\be}\sum_{\omega_l}{}^\lambda\pi_{ij}(i\Omega_n,\bK;i\omega_l,\bk)\nonumber\\
&=&\frac{1}{\be}\sum_{\omega_l}\mbox{tr} [\frac{i\omega_l+i\Omega_n+\tau_3\xi_{\bk+\bK,\lambda}-2\Delta_\lambda(\tau_1 (k_x+K_x)-\tau_2 (k_y+K_y))} {(\omega_l+\Omega_n)^2+E_{\bk+\bK,\lambda}^2}\tau_i\frac{i\omega_l+\tau_3\xi_{\bk\lambda}-2\Delta_\lambda(\tau_1 k_x-\tau_2 k_y)} {\omega_l^2+E_{\bk\lambda}^2}\tau_j]\\
&=&\frac{1}{\be}\sum_{\omega_l}\mbox{tr} [\frac{i\omega_l+i\Omega_n+\tau_3\xi_{+}-2\Delta_\lambda(\tau_1 (k_x+K_x)-\tau_2 (k_y+K_y))} {(-i\omega_l-i\Omega_n+E_{+})(i\omega_l+i\Omega_n+E_{+})}\tau_i\frac{i\omega_l+\tau_3\xi_{-}-2\Delta_\lambda(\tau_1 k_x-\tau_2 k_y)} {(-i\omega_l+E_{-})(i\omega_l+E_{-})}\tau_j]
\end{eqnarray}
where $\xi_+=\xi_{\bk+\bK,\lambda},\xi_-=\xi_{\bk\lambda}$, and the same for $E_\pm$. Then
\begin{eqnarray}
&&\frac{1}{\be}\sum_{\omega_l}{}^\lambda\pi_{00}(i\Omega_n,\bK;i\omega_l,\bk)\nonumber\\
&=&\frac{2}{\be}\sum_{\omega_l}\frac{i\omega_l(i\omega_l+i\Omega_n)+\xi_+\xi_-+4\Delta_\lambda^2(k^2+\bk\cdot\bK)} {(-i\omega_l-i\Omega_n+E_{+})(i\omega_l+i\Omega_n+E_{+})(-i\omega_l+E_{-})(i\omega_l+E_{-})}\\
&=&\frac{1}{2\be}\sum_{\omega_l}\left[(\frac{1}{-i\omega_l-i\Omega_n+E_{+}}-\frac{1}{i\omega_l+i\Omega_n+E_{+}}) (\frac{1}{-i\omega_l+E_{-}}-\frac{1}{i\omega_l+E_{-}})\right.\nonumber\\
&&\left.+(\frac{1}{-i\omega_l-i\Omega_n+E_{+}}+\frac{1}{i\omega_l+i\Omega_n+E_{+}}) (\frac{1}{-i\omega_l+E_{-}}+\frac{1}{i\omega_l+E_{-}})\frac{\xi_+\xi_-+4\Delta_\lambda^2(k^2+\bk\cdot\bK)}{E_+E_-}\right]\\
&=&-\frac{1}{2\be}\left[(1-\frac{\xi_+\xi_-+4\Delta_\lambda^2(k^2+\bk\cdot\bK)}{E_+E_-}) \sum_{\omega_l}(\frac{1}{-i\omega_l-i\Omega_n+E_{+}}\frac{1}{i\omega_l+E_{-}}+\frac{1}{i\omega_l+i\Omega_n+E_{+}} \frac{1}{-i\omega_l+E_{-}})\right.\nonumber\\
&&\left.+(1+\frac{\xi_+\xi_-+4\Delta_\lambda^2(k^2+\bk\cdot\bK)}{E_+E_-}) \sum_{\omega_l}(\frac{1}{-i\omega_l-i\Omega_n+E_{+}}\frac{1}{i\omega_l-E_{-}}+\frac{1}{i\omega_l+i\Omega_n+E_{+}} \frac{1}{-i\omega_l-E_{-}})\right]\\
&=&-\frac{1}{2}\left[(1-\frac{\xi_+\xi_-+4\Delta_\lambda^2(k^2+\bk\cdot\bK)}{E_+E_-}) (\frac{n_F(-E_-)-n_F(E_+)}{E_++E_--i\Omega_n}-\frac{n_F(E_-)-n_F(-E_+)}{E_++E_-+i\Omega_n})\right.\nonumber\\
&&\left.+(1+\frac{\xi_+\xi_-+4\Delta_\lambda^2(k^2+\bk\cdot\bK)}{E_+E_-}) (\frac{n_F(E_-)-n_F(E_+)}{E_+-E_--i\Omega_n}+\frac{n_F(E_-)-n_F(E_+)}{E_+-E_-+i\Omega_n})\right]\\
&=&-\frac{1}{2}\left[(1-\frac{\xi_+\xi_-+4\Delta_\lambda^2(k^2+\bk\cdot\bK)}{E_+E_-}) (\frac{1}{E_++E_-+i\Omega_n}+\frac{1}{E_++E_--i\Omega_n})[1-n_F(E_-)-n_F(E_+)]\right.\nonumber\\
&&\left.+(1+\frac{\xi_+\xi_-+4\Delta_\lambda^2(k^2+\bk\cdot\bK)}{E_+E_-}) (\frac{1}{E_+-E_-+i\Omega_n}+\frac{1}{E_+-E_--i\Omega_n})[n_F(E_-)-n_F(E_+)] \right].
\end{eqnarray}
Similarly,
\begin{eqnarray}
&&\frac{1}{\be}\sum_{\omega_l}{}^\lambda\pi_{11}(i\Omega_n,\bK;i\omega_l,\bk)\nonumber\\
&=&-\frac{1}{2}\left[(1-\frac{-\xi_+\xi_-+4\Delta_\lambda^2(k_x^2-k_y^2+k_xK_x-k_yK_y)}{E_+E_-}) (\frac{1}{E_++E_-+i\Omega_n}+\frac{1}{E_++E_--i\Omega_n})[1-n_F(E_-)-n_F(E_+)]\right.\nonumber\\
&&\left.+(1+\frac{-\xi_+\xi_-+4\Delta_\lambda^2(k_x^2-k_y^2+k_xK_x-k_yK_y)}{E_+E_-}) (\frac{1}{E_+-E_-+i\Omega_n}+\frac{1}{E_+-E_--i\Omega_n})[n_F(E_-)-n_F(E_+)] \right],\\
&&\frac{1}{\be}\sum_{\omega_l}{}^\lambda\pi_{22}(i\Omega_n,\bK;i\omega_l,\bk)\nonumber\\
&=&-\frac{1}{2}\left[(1-\frac{-\xi_+\xi_-+4\Delta_\lambda^2(k_y^2-k_x^2+k_yK_y-k_xK_x)}{E_+E_-}) (\frac{1}{E_++E_-+i\Omega_n}+\frac{1}{E_++E_--i\Omega_n})[1-n_F(E_-)-n_F(E_+)]\right.\nonumber\\
&&\left.+(1+\frac{-\xi_+\xi_-+4\Delta_\lambda^2(k_y^2-k_x^2+k_yK_y-k_xK_x)}{E_+E_-}) (\frac{1}{E_+-E_-+i\Omega_n}+\frac{1}{E_+-E_--i\Omega_n})[n_F(E_-)-n_F(E_+)] \right],\\
&&\frac{1}{\be}\sum_{\omega_l}{}^\lambda\pi_{33}(i\Omega_n,\bK;i\omega_l,\bk)\nonumber\\
&=&-\frac{1}{2}\left[(1-\frac{\xi_+\xi_--4\Delta_\lambda^2(k^2+\bk\cdot\bK)}{E_+E_-}) (\frac{1}{E_++E_-+i\Omega_n}+\frac{1}{E_++E_--i\Omega_n})[1-n_F(E_-)-n_F(E_+)]\right.\nonumber\\
&&\left.+(1+\frac{\xi_+\xi_--4\Delta_\lambda^2(k^2+\bk\cdot\bK)}{E_+E_-}) (\frac{1}{E_+-E_-+i\Omega_n}+\frac{1}{E_+-E_--i\Omega_n})[n_F(E_-)-n_F(E_+)] \right],
\end{eqnarray}
and
\begin{eqnarray}
&&\frac{1}{\be}\sum_{\omega_l}\left[{}^\lambda\pi_{03}(i\Omega_n,\bK;i\omega_l,\bk)+{}^\lambda\pi_{30}(i\Omega_n,\bK;i\omega_l,\bk)\right] \nonumber\\
&=&\frac{4}{\be}\sum_{\omega_l}\frac{(i\omega_l+i\Omega_n)\xi_-+i\omega_l\xi_+} {(-i\omega_l-i\Omega_n+E_{+})(i\omega_l+i\Omega_n+E_{+})(-i\omega_l+E_{-})(i\omega_l+E_{-})}\\
&=&(\frac{\xi_+}{E_+}-\frac{\xi_-}{E_-})(\frac{1}{E_++E_-+i\Omega_n}-\frac{1}{E_++E_--i\Omega_n})[1-n_F(E_-)-n_F(E_+)]\nonumber\\
&&+(\frac{\xi_+}{E_+}+\frac{\xi_-}{E_-})(\frac{1}{E_+-E_-+i\Omega_n}-\frac{1}{E_+-E_--i\Omega_n})[n_F(E_-)-n_F(E_+)],\\
&&\frac{1}{\be}\sum_{\omega_l}\left[{}^\lambda\pi_{01}(i\Omega_n,\bK;i\omega_l,\bk)+{}^\lambda\pi_{10}(i\Omega_n,\bK;i\omega_l,\bk)\right]\nonumber\\
&=&\frac{4}{\be}\sum_{\omega_l}\frac{-2\Delta_\lambda k_x(i\omega_l+i\Omega_n)-2\Delta_\lambda(k_x+K_x) i\omega_l} {(-i\omega_l-i\Omega_n+E_{+})(i\omega_l+i\Omega_n+E_{+})(-i\omega_l+E_{-})(i\omega_l+E_{-})}\\
&=&-2\Delta_\lambda(\frac{k_x+K_x}{E_+}-\frac{k_x}{E_-})(\frac{1}{E_++E_-+i\Omega_n}-\frac{1}{E_++E_--i\Omega_n})[1-n_F(E_-)-n_F(E_+)]\nonumber\\
&&-2\Delta_\lambda(\frac{k_x+K_x}{E_+}+\frac{k_x}{E_-})(\frac{1}{E_+-E_-+i\Omega_n}-\frac{1}{E_+-E_--i\Omega_n})[n_F(E_-)-n_F(E_+)],\\
&&\frac{1}{\be}\sum_{\omega_l}\left[{}^\lambda\pi_{02}(i\Omega_n,\bK;i\omega_l,\bk)+{}^\lambda\pi_{20}(i\Omega_n,\bK;i\omega_l,\bk)\right]\nonumber \\&=&\frac{4}{\be}\sum_{\omega_l}\frac{2\Delta_\lambda k_y(i\omega_l+i\Omega_n)+2\Delta_\lambda(k_y+K_y) i\omega_l} {(-i\omega_l-i\Omega_n+E_{+})(i\omega_l+i\Omega_n+E_{+})(-i\omega_l+E_{-})(i\omega_l+E_{-})}\\
&=&2\Delta_\lambda(\frac{k_y+K_y}{E_+}-\frac{k_y}{E_-})(\frac{1}{E_++E_-+i\Omega_n}-\frac{1}{E_++E_--i\Omega_n})[1-n_F(E_-)-n_F(E_+)]\nonumber\\
&&+2\Delta_\lambda(\frac{k_y+K_y}{E_+}+\frac{k_y}{E_-})(\frac{1}{E_+-E_-+i\Omega_n}-\frac{1}{E_+-E_--i\Omega_n})[n_F(E_-)-n_F(E_+)],\\
&&\frac{1}{\be}\sum_{\omega_l}\left[{}^\lambda\pi_{12}(i\Omega_n,\bK;i\omega_l,\bk)+{}^\lambda\pi_{21}(i\Omega_n,\bK;i\omega_l,\bk)\right]\nonumber \\&=&-16\Delta_\lambda^2[k_x(k_y+K_y)+k_y(k_x+K_x)]\frac{1}{\be}\sum_{\omega_l} \frac{1}{(-i\omega_l-i\Omega_n+E_{+})(i\omega_l+i\Omega_n+E_{+})(-i\omega_l+E_{-})(i\omega_l+E_{-})}\\
&=&-\frac{4\Delta_\lambda^2[k_x(k_y+K_y)+k_y(k_x+K_x)]}{E_+E_-} \left\{(\frac{1}{E_++E_-+i\Omega_n}+\frac{1}{E_++E_--i\Omega_n})[1-n_F(E_-)-n_F(E_+)]\right.\nonumber\\
&&\left.-(\frac{1}{E_+-E_-+i\Omega_n}+\frac{1}{E_+-E_--i\Omega_n})[n_F(E_-)-n_F(E_+)]\right\},\\
&&\frac{1}{\be}\sum_{\omega_l}\left[{}^\lambda\pi_{13}(i\Omega_n,\bK;i\omega_l,\bk)+{}^\lambda\pi_{31}(i\Omega_n,\bK;i\omega_l,\bk)\right]\nonumber \\&=&-8\Delta_\lambda[\xi_+k_x+\xi_-(k_x+K_x)]\frac{1}{\be}\sum_{\omega_l} \frac{1}{(-i\omega_l-i\Omega_n+E_{+})(i\omega_l+i\Omega_n+E_{+})(-i\omega_l+E_{-})(i\omega_l+E_{-})}\\
&=&-\frac{2\Delta_\lambda[\xi_+k_x+\xi_-(k_x+K_x)]}{E_+E_-} \left\{(\frac{1}{E_++E_-+i\Omega_n}+\frac{1}{E_++E_--i\Omega_n})[1-n_F(E_-)-n_F(E_+)]\right.\nonumber\\
&&\left.-(\frac{1}{E_+-E_-+i\Omega_n}+\frac{1}{E_+-E_--i\Omega_n})[n_F(E_-)-n_F(E_+)]\right\},\\
&&\frac{1}{\be}\sum_{\omega_l}\left[{}^\lambda\pi_{23}(i\Omega_n,\bK;i\omega_l,\bk)+{}^\lambda\pi_{32}(i\Omega_n,\bK;i\omega_l,\bk)\right]\nonumber \\&=&8\Delta_\lambda[\xi_+k_y+\xi_-(k_y+K_y)]\frac{1}{\be}\sum_{\omega_l} \frac{1}{(-i\omega_l-i\Omega_n+E_{+})(i\omega_l+i\Omega_n+E_{+})(-i\omega_l+E_{-})(i\omega_l+E_{-})}\\
&=&\frac{2\Delta_\lambda[\xi_+k_y+\xi_-(k_y+K_y)]}{E_+E_-} \left\{(\frac{1}{E_++E_-+i\Omega_n}+\frac{1}{E_++E_--i\Omega_n})[1-n_F(E_-)-n_F(E_+)]\right.\nonumber\\
&&\left.-(\frac{1}{E_+-E_-+i\Omega_n}+\frac{1}{E_+-E_--i\Omega_n})[n_F(E_-)-n_F(E_+)]\right\}.
\end{eqnarray}

So far we have not included long range Coulomb interaction in our calculations, and in this sense we are considering a ``neutral superconductor." At zero temperature, in the hydrodynamic limit, $\Omega_n=0, \bK\to 0$, then if $i\neq j$,
\begin{eqnarray}
\frac{1}{\be}\sum_{\omega_l}\left[{}^\lambda\pi_{ij}(0,0;i\omega_l,\bk)+{}^\lambda\pi_{ji}(0,0;i\omega_l,\bk)\right]=0\label{Eq:piij}
\end{eqnarray}
and
\begin{eqnarray}
\frac{1}{\be}\sum_{\omega_l}{}^\lambda\pi_{00}(0,0;i\omega_l,\bk)=0.\label{Eq:pi00}
\end{eqnarray}
Thus only three correlation functions are nonzero in the hydrodynamic limit: $^\lambda\Pi_{ii}(0,0)$'s for $i=1,2,3$.
\begin{eqnarray}
^\lambda\Pi_{11}(0,0)={}^\lambda\Pi_{22}(0,0) &=&-\frac{1}{2}\int\frac{d\bk}{(2\pi)^2}(1+\frac{\xi_{\bk\lambda}^2}{E_{\bk\lambda}^2})\frac{1}{E_{\bk\lambda}}\\
&=&-\frac{1}{2}N_\lambda \int_{-A}^A d\xi\frac{2\xi^2+4\Delta_\lambda^2}{(\xi^2+4\Delta_\lambda^2)^{3/2}}\\
&\approx&N_\lambda(1-2\ln\frac{A}{\Delta_\lambda}),\\
^\lambda\Pi_{33}(0,0)
&=&-N_\lambda\int_0^\infty d\xi\frac{8\Delta_\lambda^2}{(\xi^2+4\Delta_\lambda^2)^{3/2}}=-2N_\lambda,
\end{eqnarray}
where $N_\lambda$ is the density of states of band $\lambda$ at the Fermi energy, and $A$ is the cutoff. $\Delta_\lambda$ can be expressed in terms of $N_\lambda$, $A$ and $g_{\mu\lambda}$ from the gap equations, which should be derived by saddle point approximation. Combining all the results, we have the effective action for $\theta_\lambda$ in the hydrodynamic limit,
\begin{eqnarray}
S_{eff}\{\theta_\lambda\}&=&\frac{1}{8}\sum_{\Omega_n,\bK}\left\{\frac{8|g_{12}|\Delta_1\Delta_2}{g_{11}g_{22}-g_{12}^2} (\theta_1(i\Omega_n,\bK)-\theta_2(i\Omega_n,\bK))(\theta_1(-i\Omega_n,-\bK)-\theta_2(-i\Omega_n,-\bK))\right.\nonumber\\ &&\left.+\sum_{\lambda=1}^2\theta_\lambda(i\Omega,\bK)\left[2N_\lambda\left(\Omega_n^2+c_\lambda^2K^2\right)\right] \theta_\lambda(-i\Omega_n,-\bK)\right\}
\end{eqnarray}
where
\begin{eqnarray}
c_\lambda^2=\frac{n_\lambda}{mN_\lambda} +2\frac{\Delta_\lambda^2}{k_{F\lambda}^2}(1-2\ln\frac{A}{\Delta_\lambda}).
\end{eqnarray}
$c_\lambda$ is the velocity of the Bogoliubov-Anderson-Goldstone mode in band $\lambda$ for $p+ip$ superconductors. Note that the second term does not appear in $s$-wave superconductors. For quadratic dispersion, in 2D,
\begin{eqnarray}
n_\lambda=N_\lambda*\frac{1}{2}mv_{F\lambda}^2,
\end{eqnarray}
the first term is $\frac{1}{2}v_{F\lambda}^2$ (in 3D, it is $\frac{1}{3}v_{F\lambda}^2$); but with spin-orbit coupling, we do not have such a simple relation. In matrix form,
\begin{eqnarray}
S_{eff}\{\theta_\lambda\}=\frac{1}{8}\sum_{\Omega_n,\bK}[\theta_1(i\Omega_n,\bK)\ \theta_2(i\Omega_n,\bK)]\Theta^{-1}\left[\begin{array}{c} \theta_1(-i\Omega_n,-\bK)\\\theta_2(-i\Omega_n,-\bK)\end{array}\right],
\end{eqnarray}
with
\begin{eqnarray}
\Theta^{-1}&=&\left[\begin{array}{cc}2N_1\left(\Omega_n^2+c_1^2K^2\right)+B&-B\\-B&2N_2 \left(\Omega_n^2+c_2^2K^2\right)+B\end{array}\right],\\
B&=&\frac{8|g_{12}|\Delta_1\Delta_2}{g_{11}g_{22}-g_{12}^2}.
\end{eqnarray}
Solving $\det\Theta^{-1}=0$ for collective modes and making an analytical continuation $i\Omega_n\to\omega+i0$ we arrive at
\begin{eqnarray}\label{eq:omega}
\omega^2=\frac{1}{2}\left[\omega_0^2+(c_1^2+c_2^2)K^2\pm\sqrt{\omega_0^4+(c_1^2-c_2^2)^2K^4-2\omega_0^2 \frac{N_1-N_2}{N_1+N_2}(c_1^2-c_2^2)K^2}\right]
\end{eqnarray}
with
\begin{eqnarray}
\omega_0^2=\frac{N_1+N_2}{2N_1N_2}\frac{8|g_{12}|\Delta_1\Delta_2}{g_{11}g_{22}-g_{12}^2}.
\end{eqnarray}
In the limit $K\to0$,
\begin{eqnarray}
\omega^2&=&c^2K^2, \mbox{ where } c^2=\frac{N_1c_1^2+N_2c_2^2}{N_1+N_2}\ \ \mbox{for  } ``-" \mbox{ sign in front of the square root in Eq.(\ref{eq:omega})};\\
\omega^2&=&\omega_0^2+v^2K^2,\mbox{ where }v^2=\frac{N_1c_2^2+N_2c_1^2}{N_1+N_2}\ \ \mbox{for  } ``+" \mbox{ sign in front of the square root in Eq.(\ref{eq:omega})}.
\end{eqnarray}
The first solution corresponds to Bogoliubov-Anderson-Goldstone (BAG) mode, while the second to Leggett mode. Including the term proportional to $\lambda\al_R K$ does not change this result, since it contributes to correlation functions which have factors vanishing in the hydrodynamic limit, as in (\ref{Eq:piij}) and (\ref{Eq:pi00}).

In a charged superconductor, the Coulomb interaction is also treated by Hubbard-Stratonovich transformation, and a bosonic field $\varphi$ is used to decouple the four-fermion term. The effective action becomes\cite{Sharapov2002}
\begin{eqnarray}
S_{eff}\{\theta_\lambda\}=\frac{1}{8}\sum_{\Omega_n,\bK}[\theta_1(i\Omega_n,\bK)\ \theta_2(i\Omega_n,\bK)\ e\varphi(i\Omega_n,\bK)]\Theta^{-1}\left[\begin{array}{c} \theta_1(-i\Omega_n,-\bK)\\\theta_2(-i\Omega_n,-\bK)\\e\varphi(i\Omega_n,\bK)\end{array}\right],
\end{eqnarray}
where
\begin{eqnarray}
\Theta^{-1}&=&\left[\begin{array}{ccc}2N_1\left(\Omega_n^2+c_1^2K^2\right)+B&-B&-4i\Omega_nN_1\\-B&2N_2 \left(\Omega_n^2+c_2^2K^2\right)+B&-4i\Omega_nN_2\\4i\Omega_nN_1&4i\Omega_nN_2&4(2N_1+N_2+V_c^{-1}(\bK)) \end{array}\right],
\end{eqnarray}
and $V_c^{-1}(\bK)=K^2/(4\pi e^2)$, which vanishes in the hydrodynamic limit. Again solving $\det\Theta^{-1}=0$, we find
\begin{eqnarray}
\omega^2&=&\omega_0^2+v^2K^2,\mbox{ where }v^2=\frac{(N_1+N_2)c_1^2c_2^2}{N_1c_1^2+N_2c_2^2}.
\end{eqnarray}
Therefore, the gap of Leggett mode is not affected by the Coulomb interaction, while the velocity $v$ is affected. Another solution is a plasma mode,
\begin{eqnarray}
\omega^2&=&8\pi e^2(N_1c_1^2+N_2c_2^2),
\end{eqnarray}
which is just the BAG mode pushed to plasma frequency.

\end{widetext}

\end{document}